\documentclass[aps,pra,reprint,twocolumn,floatfix,notitlepage, floatfix,superscriptaddress,longbibliography]{revtex4-2}
\usepackage[utf8]{inputenc}
\usepackage{amsfonts,amsmath,amssymb,graphicx,epstopdf,verbatim}
\usepackage{bbold}
\usepackage{placeins}
\usepackage{rotating}
\usepackage{epstopdf}
\usepackage{xcolor}
\usepackage{natbib}
\usepackage[colorlinks]{hyperref}
\definecolor{darkred}{rgb}{0.5,0,0}
\definecolor{darkgreen}{rgb}{0,0.5,0}
\definecolor{darkblue}{rgb}{0,0,0.5}
\hypersetup{colorlinks,
linkcolor=darkred,
filecolor=darkgreen,
urlcolor=darkblue,
citecolor=darkgreen}
\usepackage{braket}

\newcommand{\LL}{\mathcal{L}}

\newcommand{\PP}{\mathcal{P}}

\newcommand{\rhoss}{\hat{\rho}_0}
\newcommand{\Sx}{\hat{S}_x}
\newcommand{\Sy}{\hat{S}_y}
\newcommand{\Sz}{\hat{S}_z}
\newcommand{\Jx}{\hat{J}_x}
\newcommand{\Jy}{\hat{J}_y}
\newcommand{\Jz}{\hat{J}_z}
\newcommand{\Jp}{\hat{J}^+}
\newcommand{\Jm}{\hat{J}^-}
\newcommand{\Psigma}{\hat{\sigma}}
\newcommand{\x}{\hat{\bf x}}

\newcommand{\Ham}{\hat{H}}
\newcommand{\de}{{\rm d}}
\newcommand{\nep}{{\rm e}}
\newcommand{\Ztwo}{{\mathbb Z}_2}
\newcommand{\Zn}{{\mathbb Z}_{N+1}}

\renewcommand{\Re}[1]{\mathbb{R}\mathrm{e}\left[#1\right]}

\renewcommand{\L}{\mathcal{L}}

\newcommand{\oprho}{\hat{\rho}}

\newcommand{\ph}{\varphi}
\newcommand{\dg}{\delta \Gamma}

\newcommand{\gb}{\overline{\Gamma}}

\newcommand{\Tr}[1]{\mathrm{Tr}\!\left[#1\right]}

\newcommand{\osigma}[2]{\hat{\sigma}^{#1}_{#2}}

\begin{document}

\title{Symmetries and conserved quantities of boundary time crystals\\ in generalized spin models}

\author{Giulia Piccitto }
\thanks{These two authors contributed equally to this work
\\
\href{mailto:giulia.piccitto@df.unipi.it}{giulia.piccitto@df.unipi.it} \\
\href{mailto:wauters@nbi.ku.dk}{wauters@nbi.ku.dk}
}
\affiliation{SISSA, Via Bonomea 265, I-34135 Trieste, Italy}

\author{Matteo Wauters}
\thanks{These two authors contributed equally to this work
\\
\href{mailto:giulia.piccitto@df.unipi.it}{giulia.piccitto@df.unipi.it} \\
\href{mailto:wauters@nbi.ku.dk}{wauters@nbi.ku.dk}
}
\affiliation{Niels Bohr International Academy and Center for Quantum Devices, Copenhagen University, Universitetsparken 5, 2100 Copenhagen, Denmark}
\author{Franco Nori}
\affiliation{Theoretical Quantum Physics Laboratory, RIKEN Cluster for Pioneering Research, Wakoshi, Saitama 351-0198, Japan}
\affiliation{RIKEN Center for Quantum Computing (RQC), Wakoshi, Saitama 351-0198, Japan}
\affiliation{Physics Department, University of Michigan, Ann Arbor, Michigan, 48109-1040, USA}
\author{Nathan Shammah}
\affiliation{Unitary Fund}
\affiliation{Theoretical Quantum Physics Laboratory, RIKEN Cluster for Pioneering Research, Wakoshi, Saitama 351-0198, Japan}
\affiliation{Quantum Technology Lab, Dipartimento di Fisica, Universit\`{a} degli Studi di Milano, 20133 Milano, Italy}

\begin{abstract}
We investigate how symmetries and conserved quantities relate to the occurrence of the boundary time crystal (BTC) phase in a generalized spin model with Lindblad dissipation. BTCs are a non-equilibrium phase of matter in which the system, coupled to an external environment, breaks the continuous time translational invariance.
We perform a detailed mean-field study aided by a finite-size analysis of the quantum model of a $p$,$q$-spin-interaction system, a generalized $p$-spin-interaction system, which can be implemented in fully-connected spin-$\frac{1}{2}$ ensembles.
We find the following conditions for the observation of the BTC phase: First, the BTC appears when the discrete symmetry held by the Hamiltonian, $\Ztwo$ in the considered models, is explicitly broken by the Lindblad jump operators.
Second, the system must be coupled uniformly to the same bath in order to preserve the total angular momentum during the time evolution.
If these conditions are not satisfied, any oscillatory behavior appears only as a transient in the dynamics and a time-independent stationary state is eventually  reached. 
Our results suggest that these two elements may be general requirements for the observation of a stable BTC phase relating symmetries and conserved quantities in arbitrary spin models. 
\end{abstract}

\date{\today}

\maketitle


\section{Introduction}
\label{sec:intro}

Non-equilibrium quantum many-body systems are one of the modern paradigms of research in quantum physics. 
Indeed, they represent one of the key elements to control and manipulate mesoscopic quantum systems and, therefore, they lie at the root of the growing field of quantum technologies \cite{MacFarlane_PTRSL2003,Xiang_2013_RMP,Kurizki_PNAS15,QuantumTech_NJP2018}.
Moreover they display a wide range of non-trivial dynamics \cite{break_integrability1, break_integrability2, break_integrability3, dynamics:lr1, dynamics:lr2, dynamics:lr3, entanglement:lr1, entanglement:lr4, entanglement:lr3, io1, lightcone:lr1, lightcone:lr2, lightcone:lr3} and host a plethora of interesting phenomena with no analogue in equilibrium states \cite{dqpt1, dqpt2, dqpt3, dqpt4, dqpt5,io2, io3, pretherm1, pretherm2, pretherm5}. 
Of particular interest are many-body driven-dissipative quantum systems~\cite{SiebererPRL13,Sieberer_RepProgPhys2016}, 
where dissipation and decoherence can radically change the critical properties of phase transitions~\cite{AltmanPRX15,MaghrebiPRB16} and induce an extremely rich steady-state phase diagram.

In recent years, considerable attention has been drawn by the so-called time crystals \cite{Wilczek2012,tc7, tc9}, a phase of matter characterized by a spontaneous breaking (in the thermodynamic limit) of the time-translational symmetry, that in short-ranged systems is impossible to realize at equilibrium \cite{Watanabe2015,Kozin_PRL2019}.
In their most common form, they are a sub-harmonic response to a periodic driving, thus breaking a {\em discrete} time-translational symmetry \cite{tc4,tc5,tc12}. 

More interestingly, a time crystal might appear also when there is no explicit time dependence in the Hamiltonian but the system is subjected to incoherent processes~\cite{Iemini17, tc8, Hurtado_PRL2020}. 
In this particular case, known as {\em Boundary} Time Crystals (BTC), an open system self-organizes in a time-periodic pattern, with a frequency that depends on the ratio between the model's relevant energy scales and the strength of the coupling with the environment.
From an experimental point of view, this phenomenon is very appealing because it {\em relies} on the presence of dissipation, while, usually, time-order is destroyed by the inevitable decoherence effects \cite{tc6} present in experimental setups.

BTCs have been first introduced in the context of fully connected spin models,  where permutational invariance~\cite{collective2, entanglement:lr2, classical_limit1} allows the exact numerics up to rather large systems \cite{LeePRL13,BiellaPRA15,BiellaPRA17,BiellaPRB18,MingantiSciRep16,Shammah_2018,HuybrechtsPRA19,Huybrechts19,Wauters_PRA2020} and makes the mean-field description particularly reliable \cite{collective2, collective1, collective3,collective4, collective5, equilibrium:lr3, Carollo_arxiv2020}. In contrast to discrete time crystals, however, little is still understood about the conditions needed for the emergence of the BTC phase.

An interesting topic is that of the relation between BTCs and symmetries.
The role of the symmetries in the open-system dynamics has been already under discussion \cite{Munoz_2019, Carlos2020}, especially in the case of open systems with decoherence-free subspaces~\cite{Baumgartner_2008_NJP,Buca_2012,Albert_2014_PRA,Kockum_2018, Manzano_AdvPhys2018,  Munoz_2019,Minganti_2020_arXiv,Lieu_2020}.
Although BTCs share some phenomenology with them, they have different physical roots.
The BTC phase, indeed, is well defined only in the thermodynamic limit, being associated with the spontaneous symmetry breaking of the time translational invariance. 
On the other hand, if the dynamics has decoherence-free subspaces, a non-stationary behavior may emerge after an initial relaxation at any system size.
In particular, symmetry-preserving dissipation concurs to split the non-decaying part of the Hilbert space into disjoint sectors, leading to permanent oscillations with a frequency directly inherited from the Hamiltonian \cite{Buca_2019_NatureComm}.
In BTCs, instead, the frequency depends on a nontrivial interplay between the parent closed system dynamics and the incoherent dissipation or driving.


In Ref.~\cite{Iemini17}, it has been shown that both a free spin in a static magnetic field and the fully connected Ising model exhibit a transition from a normal phase to a time-ordered one, 
driven by the system-environment coupling strength, when all spins coupled to the same bath through a collective~\cite{nota1} operator. 
However, an apparent contradiction  emerged in Ref.~\cite{Wang2020}, where no evidence of time-ordering was found in the closely related $p$-spin model, where $p$ refers to the order of the highest interaction in the Hamiltonian, $p\geq 2$.
 
In this work we solve this seeming contrast and clarify the mechanism behind the onset of persistent oscillations in the expectation values of physical observables.
To this purpose, we consider a generalized $p$-spin Hamiltonian, coupled to the environment via Lindblad operators -- the model can be implemented in a fully-connected spin-$\frac{1}{2}$ ensemble with collective dissipation, analogously to Refs.~\cite{Iemini17, Wang2020}. 
The dissipative dynamics is studied by combining a mean-field approach and an exact finite-size analysis,
where we solve directly the Lindblad master equation~\cite{lindblad1, lindblad3} 
for the density matrix $\dot{\oprho}=\LL[\oprho]$, where the evolution is determined by the Liouvillian superoperator $\LL$. 
In this setting, the BTC phase is manifested by the spontaneous breaking of the time-translational invariance $\nep^{\LL t}\oprho_0 = \oprho_0$.
In other words, in the thermodynamic limit the system does not reach a stationary state, as it does, instead, for finite sizes.

We find that BTCs survive when two elements are simultaneously present. 
First, the Hamiltonian part of the model possesses a discrete symmetry, $\Ztwo$ for the collective spin-$\frac{1}{2}$ system, which must be explicitly broken by the operators coupling the system with the environment. 
The model used in Refs.~\cite{Wang2020}, contrarily to Ref.~\cite{Iemini17}, does not satisfy this symmetry requirement, hence explaining the absence of BTCs.

Second, the system must have a strong symmetry, namely an observable that commutes both with the Hamiltonian and the Lindblad operators. This requirement is fulfilled by including in the open-system dynamics only collective incoherent processes, as was guessed in Refs.~\cite{Shammah_2018,tc8}, meaning that all sites need to be coupled uniformly to global baths. 
Physically, this is related to the presence of a conserved quantity, which in the considered spin models is the total angular momentum.

In the absence of either one of the two conditions, we find that the system always relaxes towards a stationary state, time-translational invariance is recovered in the long-time limit and all information on the initial condition is lost. 
When both are satisfied, instead, the boundary time crystal phase is present; at the mean-field level, it is identified by the presence of closed periodic trajectories in the semiclassical phase space, 
while in finite sizes it is revealed by oscillations with a damping rate diverging in the thermodynamic limit. Supported by numerical evidence in a study including dissipation to ever increasing bath sizes -- ranging from local to collective--, we argue that these elements may be necessary conditions for the existence of BTCs in general spin systems, beyond homogeneous fully-connected spin models.

Interestingly, also the finite-size density matrix $\rhoss$ of the stationary state can identify the BTC phase: when time-order is present in the thermodynamic limit, $\rhoss$ cannot be approximated by a well-defined mean-field ansatz, but rather it is the average over all the states on the time-crystal trajectory~\cite{tcsavona}.

The article is organized as follows: In Sec.~\ref{sec:model} we introduce the model we investigate and its mean-field description. 
In Sec.~\ref{Sec:BTCs} we review some of the main results known in the literature within the more general framework we present in this paper.
Our main results are presented in Sec.~\ref{sec:symmetry}, where we focus on the symmetry requirements to the BTC phase, and in Sec.~\ref{sec:colldecay}, where we show the effects of non-collective Lindblad operators.
In Sec.\ref{sec:experiments} we discuss possible experimental realizations of boundary time-crystals. Finally, we summarize our findings in Sec.~\ref{sec:conclusions}.

\section{Theoretical framework}
\label{sec:model}
We consider a generalization of the well know $p$-interacting spin model~\cite{Gardner_NPB85,Filippone_PRA2011,Bapst12,Wauters_PRA2017} described by the Hamiltonian ($\hbar = 1$)
\begin{equation}
\begin{aligned}\label{Eq:Hamiltonian_generic}
    \Ham &= -N \left(\omega_z \Jz^p + \omega_x \Jx^q \right),
\end{aligned}
\end{equation}
with $p, q \in \mathbb{N}$ and the spin algebra is given by $[\hat{J}_x,\hat{J}_y]=i2\hat{J}_z/N$, $[\hat{J}_+,\hat{J}_-]=2\hat{J}_z/N$, where,
without loss of generality, we consider these as collective spins from a system of $N$ all-to-all interacting spin-$\frac{1}{2}$. $\hat{J}_\alpha= \sum_i \Psigma^\alpha_i/N$  are the (collective) magnetization operators and $\Psigma_i^\alpha$ are the Pauli matrices acting on the $i$-th site. 
The dynamics we consider is given by the Lindblad master equation~\cite{lindblad1,lindblad2} 
\begin{equation}\label{Eq:master}
	\begin{aligned}
		\frac{\de}{\de t}{\oprho} = - &i [\Ham, \oprho] + N\Gamma^\uparrow \left(\Jp \oprho \Jm - \frac{1}{2}\{\Jm \Jp, \oprho\} \right)\\
		+ &N\Gamma^\downarrow \left(\Jm \oprho \Jp - \frac{1}{2} \{\Jp \Jm, \oprho\} \right) \ ,\\
	\end{aligned}
\end{equation}
where $\hat{J}^\pm = \Jx \pm i \Jy$ describe two collective incoherent processes with associated rates $\Gamma^{\uparrow (\downarrow)}$.
Note that Eq.~(\ref{Eq:master}) in general couples the spin ensemble to two separate baths, as the $\Gamma^{\uparrow (\downarrow)}$ rates can be independently chosen. This includes the case of a single bath at detailed balance, in which $\Gamma^{\uparrow }=\Gamma_0\left(1+n_T\right)$, and $\Gamma^{\downarrow }=\Gamma_0 n_T$ due to the favored spin alignment in the upward direction of Eq.~(\ref{Eq:Hamiltonian_generic}), with $n_T$ the thermal occupation number, which goes to zero at $T=0$, and $\Gamma_0$ a coefficient fixed for the considered model.
Depending on the largest between $\Gamma^\downarrow$ and $\Gamma^\uparrow$, the Lindblad operators favor spin alignment along the upward ($\Jp$) or the downward ($\Jm$) $z$ direction, explicitly breaking the $\Ztwo$ symmetry the Hamiltonian displays for $p$ even. 

Notice that, since $[\hat H, \hat{J^2}]=[\hat J_-, \hat{J^2}]=[\hat J_+, \hat{J^2}]=0$, the total angular momentum $\hat{J^2}$ is a \emph{strong symmetry} \cite{Buca_2012,Albert_2014_PRA} of the set of generalized spin models considered and Eq.~\eqref{Eq:master} preserves the total angular momentum $\hat{J^2}$\footnote{We want to remark that Noether theorem can not be generalized at the operator level in the case of open quantum systems. For this reason there is no one-to-one correspondence between symmetries and conserved quantities. In particular, it is possible to find conserved quantities that are not associated to any symmetry, but any strong symmetry in the model has an associated conserved quantity\cite{Albert_2014_PRA}.\label{fn:sym-cc}}.
Moreover, besides for the treatment of local baths performed in Sec.~\ref{sec:colldecay}, hereafter we can generalize the models beyond their derivation in terms of collections of spin-$\frac{1}{2}$, and consider them genuine $(N+1)$-spins, further opening up quantum simulation experimental possibilities, as detailed in Sec.~\ref{sec:experiments}. 

The mean-field equations of motion for the expectation values of the magnetization operators $X, Y, Z = \braket{\hat{J}_x},\braket{\hat{J}_y},\braket{\hat{J}_z}$ are obtained with a Gutzwiller approximation, $\braket{\hat{J}_\alpha \hat{J}_\beta} = \braket{\hat{J}_\alpha} \braket{\hat{J}_\beta}$, where $\alpha, \beta = x,y,z$ and $\braket{\cdot} = \text{Tr} [\hat\rho\  \cdot]$ is the expectation value on the state $\hat\rho$.
 With this approximation, the equations for the macroscopic variables are
\begin{equation}
\left\{
\begin{aligned}
	& \dot{X} = 2p \omega_z  Z^{p-1}Y - 2\delta \Gamma ZX , \\
	& \dot{Y} = 2XZ \left (q\omega_x X^{q-2}- p\omega_z Z^{p-2} \right) - 2\dg ZY , \\
	& \dot{Z} = -2q\omega_x Y X^{q-1} + 2 \dg (1 - Z^2),
	\label{Eq:dynamics}
\end{aligned}
\right.
\end{equation}
with $\delta \Gamma = \Gamma^\uparrow - \Gamma^\downarrow$ (the full derivation can be found in App.~\ref{App:A1}).

We introduce in notation $\dot{\bf R}=(\dot{X},\dot{Y},\dot{Z})$ and study the solutions of this system for $\dot{\bf R}=0$, i.e. ${\bf R}=(X_\text{st}, Y_\text{st}, Z_\text{st})$ are the stationary states of the dynamics. In what follows, we refer to a \emph{ferromagnetic} stationary state if $|Z_\text{st}| > 0$ and to a \emph{paramagnetic} one if $Z_\text{st}= 0$.
The real part of the eigenvalues of the associated Jacobian, i.e. the real part of the Lyapunov exponents, provides a classification of the stationary states: if all the eigenvalues have positive (negative) real part, the trajectories are attracted toward (repelled from) the relative state. 
More interestingly, if the real part is zero and the Lyapunov exponents are purely imaginary \footnote{The presence of purely imaginary Lyapunov exponents is necessary for the existence of the BTCs but it is not sufficient. For instance, it is possible to have purely imaginary Lyapunov exponents associated to slow spiraling trajectories in which the damping of the oscillations is a non-linear effect of the dynamics.} we are in presence of a marginal fixed point generating periodic orbits, and the stationary state is associated with a BTC trajectory \footnote{Notice that in the BTC phase every initial condition is associated to a different periodic orbit that keeps memory of the evolution of the initial state. This a substantial difference with the dynamics in presence of limit cycles in which all the trajectories eventually reach the same periodic orbit independently on the initial conditions.}.

When studying a finite size system, a fundamental role is played by the structure of the Liouvillian superoperator~\cite{lindblad2, oqs1, oqs2, oqs_daley, lmg:oqs2, lmg:oqs3} $\LL$, defined through Eq.~\eqref{Eq:master} as $\dot{\oprho}=\LL[\rho]$.
For a finite system size $N$, in general, the spectrum of $\LL$ is gapped and the dynamic relaxes towards a stationary state $\rhoss$, defined as the right eigenvector of $\L$ with eigenvalue $\lambda_{0}=0$.
The other eigenvectors, instead, are associated with eigenvalues $\lambda_{i \neq 0}$ with $\Re{\lambda_{i\neq 0}} <0$ and give information on the transient dynamics. For instance, the real part of the first nonzero eigenvalue $|\Re{\lambda_1}|$, the Liouvillian gap, describes the relaxation rate.

The BTC phase, instead, is characterized by the presence of at least one complex eigenvalue (and its complex conjugate) 
whose real part vanishes in the thermodynamic limit, while the imaginary part saturates to a constant value \cite{Russomanno17, tcsavona}. 
In this case, the real and the imaginary parts of the eigenvalues are associated with the damping of the oscillations in finite systems and to their frequency, respectively. 

\section{Boundary time crystals in $p,q$-interacting spin models: An overview}\label{Sec:BTCs}
To exemplify our findings, we start with reviewing the results known in the literature: the
 Ising Hamiltonian with $z$--interactions ($p = 2$ and $q =1$), and the Ising Hamiltonian with $x$--interactions ($p = 1$ and $q = 2$), both in the presence of dissipation, which have been studied in Refs.~\cite{Iemini17} and~\cite{Wang2020}, respectively.
Here we want to summarize the main features that will be essential for what comes next. 

\subsection{Free spins ($p =1, q = 1$)}
Before proceeding let us briefly comment on the case of free spins in a magnetic field lying in the $x$-$z$-plane. 
The Hamiltonian reads ($p =1, q = 1$ in Eq. \eqref{Eq:Hamiltonian_generic})
\begin{equation}\label{Eq:Hamiltonian_free}
    \Ham^\text{free} = -N \left(\omega_z \Jz + \omega_x \Jx \right).
\end{equation}
Without loss of generality, we take $\omega_x, \ \omega_z \geq 0$ throughout the whole paper.

The limit $\omega_x = 0$ is trivial: we have two possible ferromagnetic stationary solutions  $ {\bf R} = \left( 0, \ 0, \ \pm 1 \right)$, obtained by solving $\dot{\bf R}=0$; one of them is stable and the other is unstable, depending on the sign of $\dg$. 
When $\omega_z = 0$, as discussed in \cite{Iemini17}, the system has two different phases divided by the critical value of the dissipation $\dg_c = \omega_x$: for $\dg < \dg_c $ the system is in a BTCs phase, while for $\dg > \dg_c$ the system again has two stationary states ${\bf R} = \left( 0, \ 0, \ \pm 1 \right)$.
As soon as $\omega_z > 0$, the system falls inevitably in the ferromagnetic phase, where the time crystal order is destroyed and there are only trivial solutions with magnetization $Z = \omega_z/\sqrt{\omega_x^2 + \omega_z^2 - \dg^2}$.

As a side remark, we would like to comment on the analogies between the mean-field limit in Eq.~\eqref{Eq:dynamics}, for $p=q=1$, and the Lorenz equations~\cite{Lorenz_JAtmSci63}, which are one of the best known examples of deterministic chaos in classical systems. 
Despite this similarity, our model we cannot have chaotic behavior because of the conservation of angular momentum $X^2+Y^2+Z^2=1$, due to the collective jump operators, that reduces the effective dimension of the phase space. 
Chaotic dynamics, however, might be recovered by introducing an explicit time dependence in the equations of motion. 

\subsection{$z$--interactions ($p =2, q = 1$)}\label{Subsec:z}
Let us now move to the fully connected Ising chain in transverse field, with interactions along the $z$ direction described by the Hamiltonian ($p =2, q = 1$ in Eq.~\eqref{Eq:Hamiltonian_generic})
\begin{equation}
\begin{aligned}\label{Eq:Hamiltonian_z}
    \Ham^{(z)} &= -N \left(\omega_z \Jz^2 + \omega_x \Jx\right).
\end{aligned}
\end{equation}
\begin{figure}[h!]
    \includegraphics[width = 0.5\textwidth]{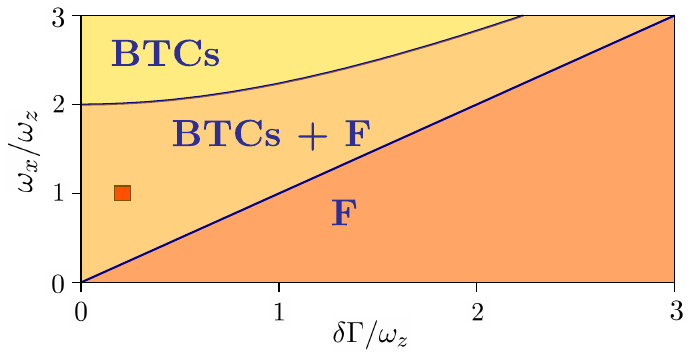}
    \caption{Phase diagram of the system in Eq.~\eqref{Eq:Hamiltonian_z}, with $z$-interactions and $x$-transverse field in the Hamiltonian and collective dissipation greater than collective incoherent pumping, $\dg>0$ . We plot the class of solutions as a function of the transverse field $\omega_x$ and the dissipation rate $\dg$ both in units of $\omega_z$. F means that we have ferromagnetic solutions only, BTCs that we have marginal solutions only, F + BTCs means that both phases coexist. The red square marks the parameters used in Fig. \ref{Fig:pp_z}.}
    \label{Fig:ddf}
\end{figure}
Without loss of generality, let us assume $\omega_x > 0$. 
The mean-field equations in Eq.~\eqref{Eq:dynamics} have four fixed points, two ferromagnetic solutions, and two paramagnetic ones. 

The ferromagnetic solutions are given by
\begin{equation}
    {\bf F_\pm} = \left(\frac{2\omega_x\omega_z}{4\omega_z^2+\dg^2}, \  \frac{\dg \omega_x}{4\omega_z^2 + \dg^2}, \  \pm\sqrt{1 - \frac{\omega_x^2}{4\omega_z^2 + \dg^2}} \right),
\end{equation}
one of which is stable and the other unstable, in fact the system is attracted toward the positive or negative magnetization state depending on the sign of $\dg$.
From the constraint $X^2+Y^2+Z^2=1$ we deduce the existence condition for these solutions $\omega_x < \omega_{x_c} = \sqrt{4\omega_z^2 + \dg^2}$. 

The two paramagnetic solutions are
\begin{equation}
    {\bf P_\pm} = \left( \pm \sqrt{1-\frac{\dg^2}{\omega_x^2}}, \   \frac{\dg}{\omega_x}, \  0\right) ,
\end{equation}
which exist only when the dissipation rate is smaller than the transverse field $\dg \le \omega_x$. 
Among the two, the most interesting is the negative one, since it is the marginal point that acts as the generator of the periodic orbits characterizing the BTC phase.

Having gathered this information, we can draw the phase diagram in Fig.~\ref{Fig:ddf}, in which we show the phases of the system as a function of $\omega_x$ and $\dg$, both in unit of $\omega_z$: the system has a ferromagnetic phase (F) for $\dg > \omega_x$, a BTCs phase for $\omega_x > \sqrt{4 \omega_z^2 + \dg}$ and a region where they coexist. Note that values of $\dg > \omega_z$ are attainable in driven systems whose interaction-picture Hamiltonian takes the form of Eq.~\eqref{Eq:dynamics}, even if dissipation mechanisms are perturbative effects \cite{Dimer07}.
In the coexistence phase the dynamics depends on the initial conditions: the closer the parameters $\omega_x$ and $\delta\Gamma$ are to the purely ferromagnetic phase, the larger is the basin of attraction for the corresponding fixed point. 
In the opposite situation, close to the boundary with the BTC phase, most of the initial conditions lead to periodic trajectories while a small number of them relaxes towards the ferromagnetic point.

\begin{figure}
    \includegraphics[width = 0.5\textwidth]{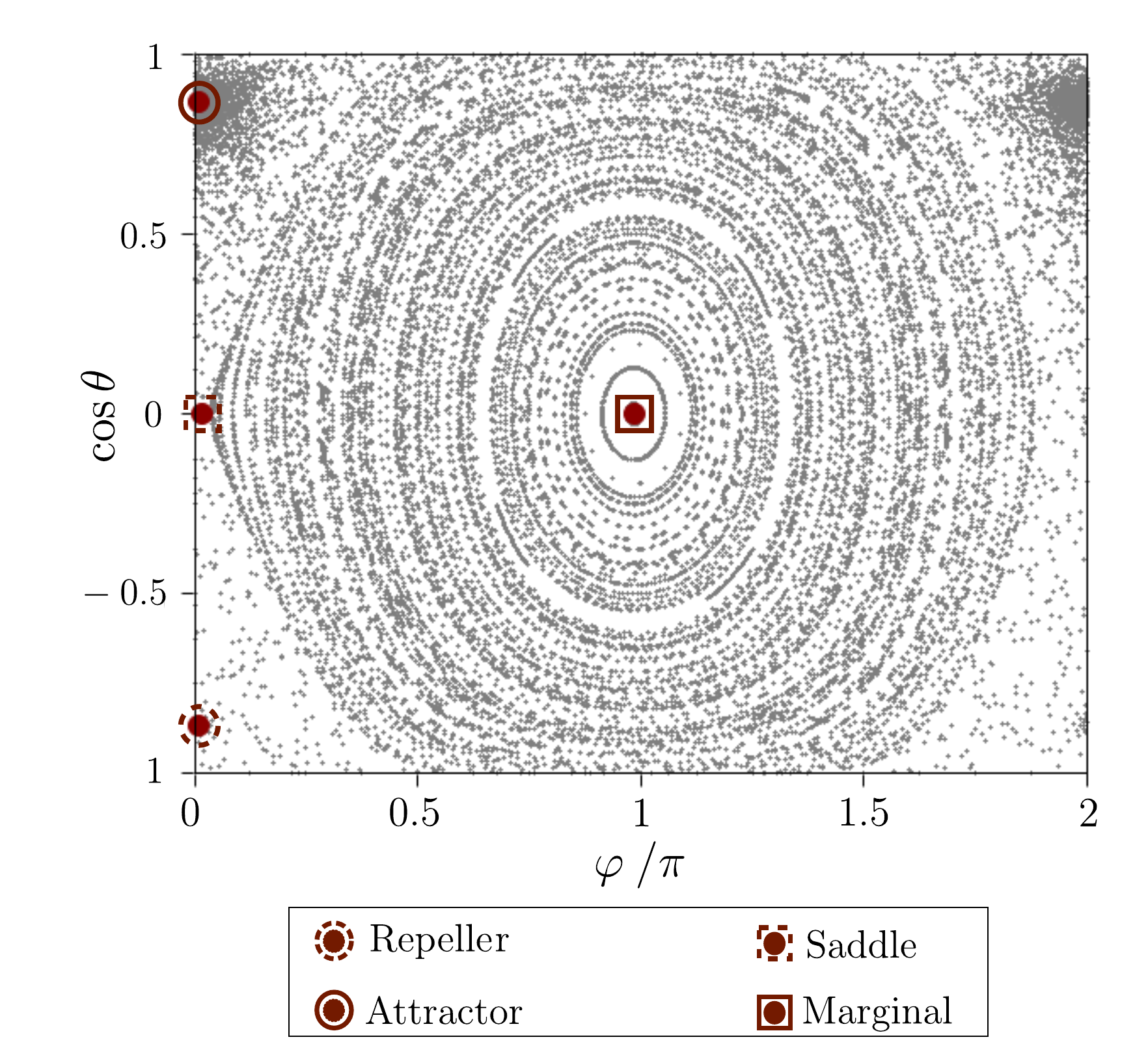}
    \caption{Phase portrait of the dynamical system in Eq.~\eqref{Eq:dynamics} for $p = 2$ and $q = 1$. The parameters $\omega_z = \omega_x = 1$, $\delta\Gamma=0.2\omega_x$ are chosen to be in the coexistence region. The black lines are the trajectories associated to different initial conditions, while the red dots are the stationary states of the system, two ferromagnetic solutions and two paramagnetic ones. Of the two ferromagnetic solutions, the positive one (solid circle) is stable and attracts the trajectories, while the negative one (dashed circle) is unstable and therefore repels them. The negative paramagnetic solutions (solid square) generate periodic orbits associated to the BTC behavior, while the positive paramagnetic one (dashed square) separates the ferromagnetic basin of attraction from the BTC one.}
    \label{Fig:pp_z}
\end{figure}

A more intuitive way to visualize this information is by looking at the phase portrait obtained by studying the dynamics in polar coordinates, which correspond to the parametrization
\begin{equation}
    {\bf R} =  \left(\sin \theta \cos \varphi, \ \sin \theta \sin \varphi, \ \cos \theta \right),
    \label{Eq:pp_z}
\end{equation}
see appendix \ref{App:A2} for more details on the derivation.

In Fig. \ref{Fig:pp_z} we show the phase portrait for the phase coexistence region, with $\omega_x = \omega_z$ and $\dg = 0.2\omega_z$ as a function of the polar angle $\ph$ and its conjugate variable $\cos \theta$, with $\theta$ the azimuthal angle. 
The red dots are the solutions of the system in Eq.~\eqref{Eq:dynamics}, with the left-hand side equal to zero, while the grey lines are trajectories associated with different initial conditions.
Each fixed point clearly influences the trajectories in a different way:
the ferromagnetic state with positive (solid circle) magnetization  ($\cos\theta >0$) is stable and therefore attracts the nearby trajectories, while the negative(dashed circle) one repels them.
The marginal paramagnetic (solid square) solution at $(\ph = \pi, \cos\theta = 0)$, instead, generates the periodic orbits associated with the time crystal behavior. 
Finally, there is a second paramagnetic ``saddle'' (dashed square) fixed point at $(\ph = 0, \cos\theta = 0)$ which separates the region of influence of all other stationary solutions.

\subsection{$x$--interactions ($p =1, q = 2$)}\label{Subsec:x}

Let us now switch the direction of the interaction and the transverse field in the Hamiltonian, i.e. we consider a fully connected Ising chain in transverse field with interactions along the $\x$ direction, described by ($p =1, q = 2$ in Eq.~\eqref{Eq:Hamiltonian_generic}):
\begin{equation}
\begin{aligned}\label{Eq:Hamiltonian_x}
    \Ham^{(x)} &= -N \left(\omega_x \Jx^2 + \omega_z \Jz\right) \ .
\end{aligned}
\end{equation}

\begin{figure}
    \includegraphics[width = 0.49\textwidth]{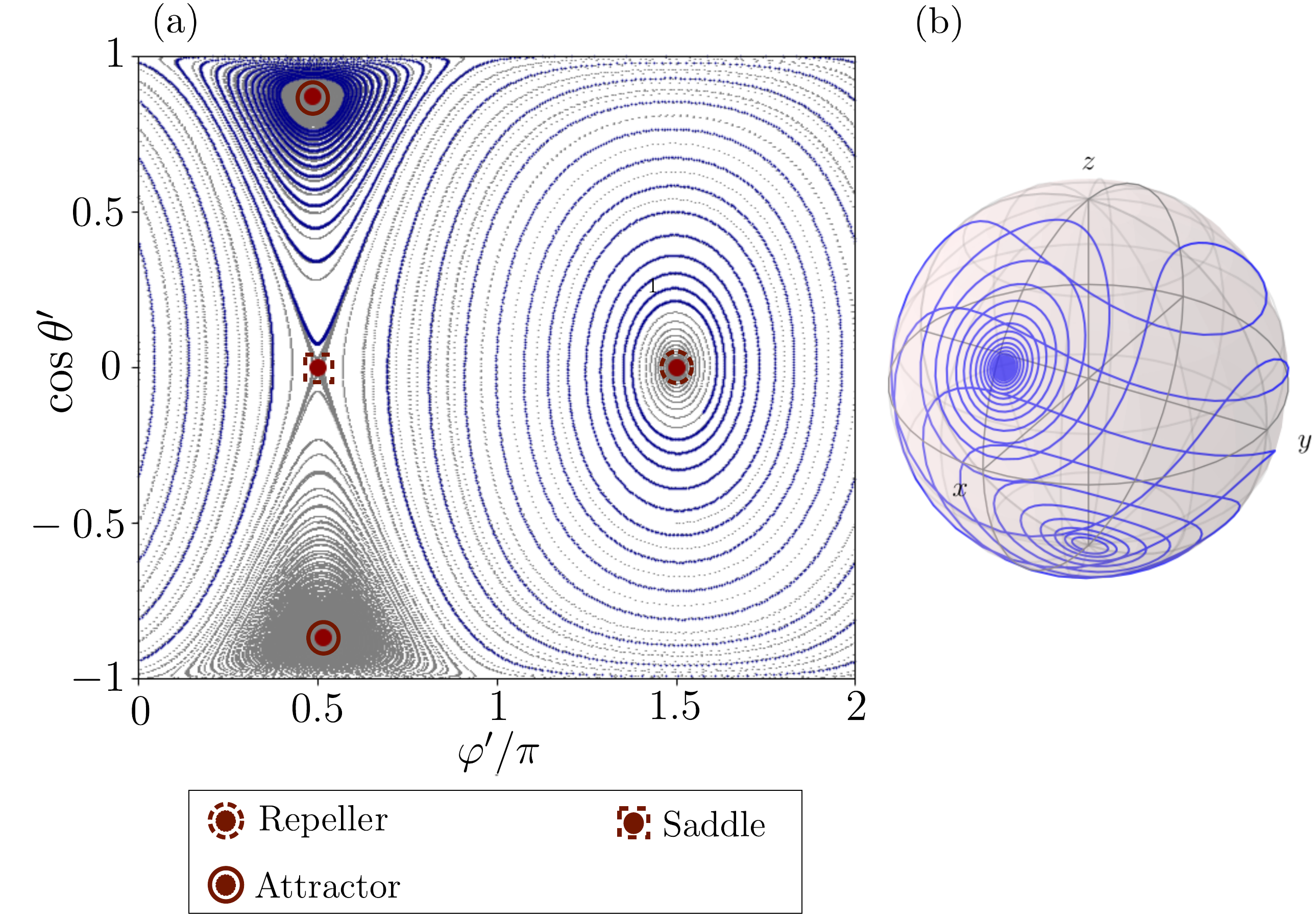}
    \caption{Panel (a): Phase portrait of the dynamical system in Eq.~\eqref{Eq:dynamics} for $p = 1$ and $q = 2$. The parameters $\omega_z = \omega_x$, $\delta\Gamma=0.2\omega_x$ correspond to a ferromagnetic phase. The black lines are the trajectories associated to different initial conditions, while the red dots are the stationary states of the system, two x-ferromagnetic ($|X| > 0$) solutions (solid circles) and two x-paramagnetic ($X = 0$) ones. We notice that the x-ferromagnetic solutions are both stable, reflecting the $\Ztwo$ symmetry of the Hamiltonian. The negative x-paramagnetic solution (dashed circle) is a repeller, while the other one (dashed square) separates the positive x-ferromagnetic basin of attraction from the negative x-ferromagnetic one.
    The Bloch sphere in panel (b) shows a particular trajectory which well represent the behavior close to attractive and repulsive fixed points. The same trajectory is depicted with a continuous blue line also in the main panel.}
    \label{Fig:pp_x}
\end{figure}

Although the closed systems described by Eqs.~\eqref{Eq:Hamiltonian_z} and \eqref{Eq:Hamiltonian_x} are identical, 
the specific form of the Lindblad operators leads to very different dynamics when coupling to the environment.
In this case, there is a single line in the phase diagram with time crystal order, corresponding to $\omega_z = 0$: as soon as $\omega_z > 0$ BTCs are destroyed and the system relaxes toward a stable fixed point, that can be either x-paramagnetic ($X=0$), or x-ferromagnetic ($|X|>0$).
In Fig. \ref{Fig:pp_x} we show the phase portrait for $\omega_z = \omega_x = 1$ and $\dg = 0.2 \omega_x$. 
In order to avoid the singularities at the poles of the unit sphere, here we use a different parametrization with respect to standard polar coordinates
\begin{equation}
       {\bf R} = \left(\cos \theta', \ \sin \theta' \cos \varphi', \ \sin \theta \sin \varphi' \right) \,
       \label{Eq:pp_x}
\end{equation}
where $\ph'$ and $\theta'$ are the polar and azimuthal angle defined with respect to the $x$-axis.
Looking at the phase portrait in panel (a) of Fig. \ref{Fig:pp_x} we identify two attractive x-ferromagnetic states (solid circles), reflecting the $\Ztwo$ symmetry of the Hamiltonian, and one repelling x-paramagnetic state (dashed circle) with $X = 0$ and $Z= -1$.

Although, at first sight, it might appear that this point is surrounded by closed orbits, actually the trajectories are escaping from the fixed point, describing a \emph{spiral}. This is more evident by looking at the behavior on the Bloch sphere in panel (b) of Fig. \ref{Fig:pp_x} of the corresponding blue trajectory in the phase space.
Thus, in contrast with the previous situation, there are no periodic orbits associated with a time-crystalline phase. 
This holds for any nonzero value of $\omega_z$: the stronger the transverse field, the faster the trajectories will collapse.
Finally, the last x-paramagnetic solution (dashed square) separates the basin of attraction of the positive x-ferromagnetic solution to that of the negative x-ferroamagnetic one.

\section{BTCs existence condition 1: Symmetry}\label{sec:symmetry}
In this section, we present the mean-field and exact finite-$N$ dynamics of the generalized $p$-spin model. 
From our results, it emerges that to observe BTCs the Hamiltonian needs to be $\Ztwo$ invariant and the Lindblad operators must break explicitly this symmetry.
This implies that the BTC phase exists only for even values of $p$, while $q$ can be either even or odd.

Although in closed system there is a perfect correspondence when exchanging $p \longleftrightarrow q$ and $\omega_z \longleftrightarrow \omega_x$, this is no longer true in the presence of dissipation.
In the case of even $p$, the $\Ztwo$ symmetry is generated by $G_z=\prod_j \osigma{x}{j}$. This operator commutes with the Hamiltonian but acts non-trivially on the jump operators $G_z \hat{J}^\pm G^\dagger_z = \hat{J}^\mp$. This action corresponds to an effective switch of $\Gamma^\uparrow$ and $\Gamma^\downarrow$ and, consequently, to an effective switch of the preferred alignment of the spins, breaking the $\Ztwo$ symmetry of the system.
In the case of even $q$, instead, the symmetry generator is $G_x=\prod_j \osigma{z}{j}$ and its action on the jump operators is $G_x \hat{J}^\pm G^\dagger_x = -\hat{J}^\pm$, hence $G_x$ does not commute with $\hat{J}^\pm$. Despite this, we observe that, differently from the previous case, the Lindblad equation remains unchanged under the effect of the symmetry operator (the jump operators always come in pairs and the minus signs cancel), leaving a {\em weak} $\Ztwo$ symmetry in the open system.

\begin{figure}[ht!]
     \includegraphics[width = 0.49\textwidth]{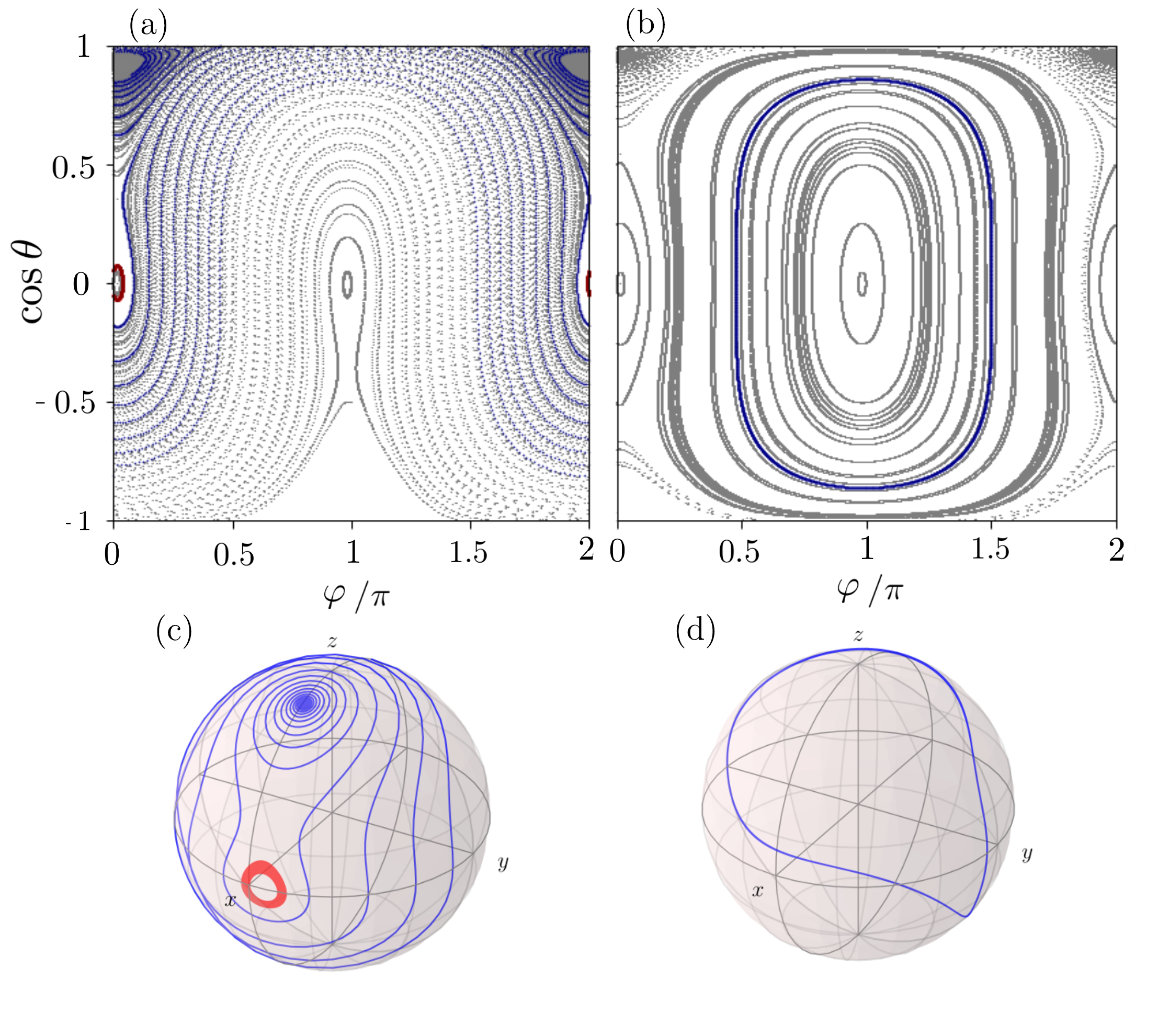}
      \caption{Top panels: Phase portraits obtained by solving Eq.~\eqref{Eq:dynamics} with $q = 1$, $p = 3$ (a) and $p = 4$ (b), fixed $\omega_z = \omega_x$ and $\delta\Gamma=0.2\omega_z$. The red trajectories correspond to the initial condition $\ph_0 = 0.10$ and $\theta_0 = 1.47$. Both diagrams present two ferromagnetic solutions and two paramagnetic ones. The ferromagnetic solutions are in both cases one attractive and the other repelling trajectories. The nature of the paramagnetic solutions, instead, depends on the value of $p$. In fact, only for $p = 4$ (d), namely the $\Ztwo$-symmetric case, we have a marginal fixed point generating periodic orbits, while for $p = 3$ (c) it slowly attracts the trajectories. 
      Bottom panels: typical trajectories depicted on the Bloch sphere for $p=3$ (c) and $p=4$ (d); the same trajectories are highlighted with the same color also in the phase portraits in (a) and (b).
      The ``thickness'' of the red trajectory in (c) signals that it is slowly spiralling towards the $X=1$ fixed point.
      }
    \label{Fig:pp_p3_p4}
\end{figure}
\subsection{Mean-field analysis}\label{Subsec:mf}
Here, we explore the mean-field dynamics of the generalized $p$-spin model in Eq.~\eqref{Eq:Hamiltonian_generic} to discuss the symmetry conditions necessary to observe BTCs. 
As explained in Sec. \ref{sec:model}, the Hamiltonian in Eq. \eqref{Eq:Hamiltonian_generic} displays a $\Ztwo$ symmetry on the $z$($x$) direction for even values of $p$($q$).
The Lindblad operators $\Jp$ and $\Jm$ favor the positive or negative alignment of the spins along the $z$ direction depending on the sign of $\delta \Gamma=\Gamma^{\uparrow}-\Gamma^{\downarrow}$. 
Our claim is that, for such a dissipation, the BTC phase appears only in the presence of $\Ztwo$ symmetry along the $z$ direction.
Hence, we expect to observe BTCs  for even $p$, independently on the value of $q$.

To prove this, we present two different scenarios: First we consider no symmetries along the $x$ directions ($q=1$) and we show that BTCs arise only for even $p$; then we fix $p = 2$ and we observe that BTCs exist for any value of $q$.

The more intuitive way to do so is by looking at the phase portrait for different values of $p$ and $q$. From now on we will parametrize the sphere as in Eq.~\eqref{Eq:pp_z}.
All the results (unless specified) are obtained with $\omega_x = \omega_z $ and $\dg = 0.2\omega_z$.

First, we fixed $q = 1$. 
In the top panels of Fig.~\ref{Fig:pp_p3_p4} we show the phase portrait for $p = 3$ (a) and $p = 4$ (b), a non-symmetric and to a $\Ztwo$-symmetric Hamiltonian, respectively. 
At first sight, the two figures seem to share the same physics: 
the trajectories are either attracted toward a positive ferromagnetic stationary state or stuck into periodic orbits. 
\begin{figure*}[ht!]
    \includegraphics[width = \textwidth]{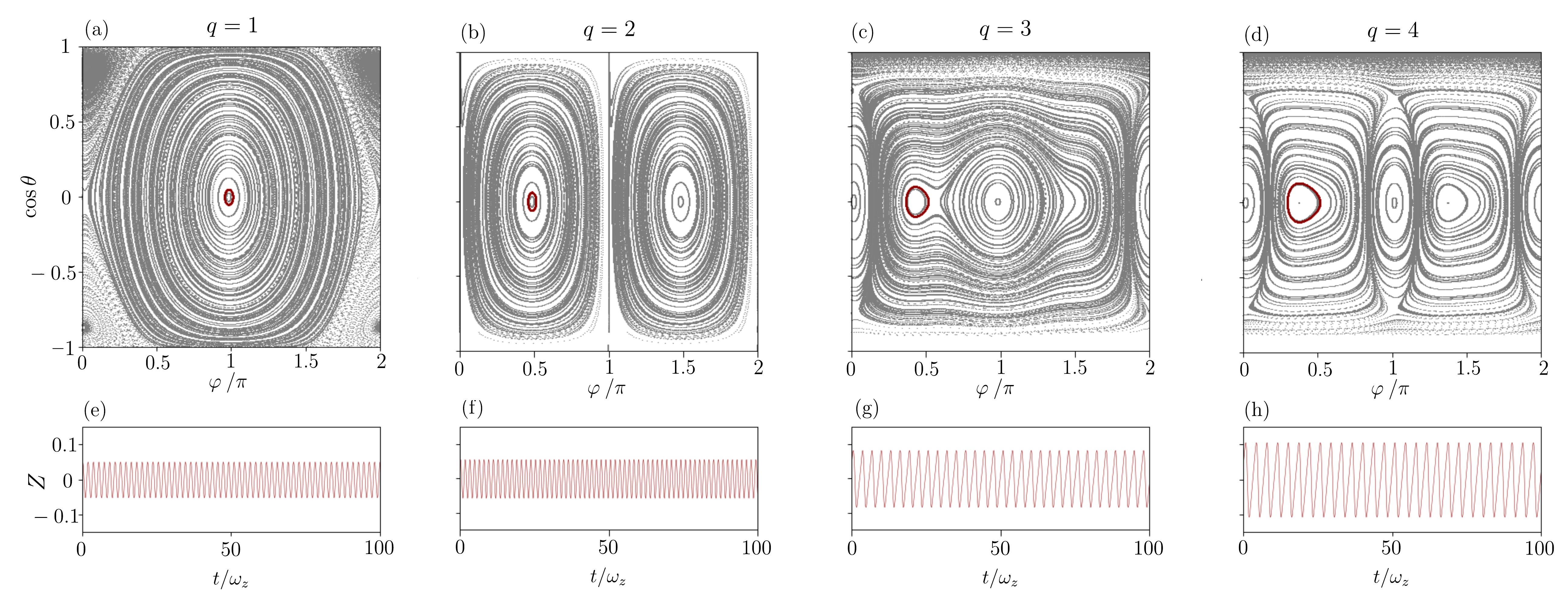}
      \caption{Top panels: phase portraits for $p = 2$ and $q = 1$ (a), $q = 2$ (b), $q =3$ (c), $q = 4$ (d). We set $\omega_x = \omega_z $, $\delta\Gamma=0.2\omega_z$. For all the values of $q$, we observe periodic orbits, indicating BTCs. The blue and orange squares in panel (a) Fig. \ref{Fig:pp_p2_q} mark the parameters used in Fig. \ref{fig:exact_diff}. The bottom panels show the behavior in time of the longitudinal magnetization evaluated over the red trajectories in the relative phase portrait (corresponding to the initial conditions $\ph_0 = 3.10$, $\theta_0 = 1.47$ for $q = 1$ (e) and  $\ph_0 = 1.57$, $\theta_0 = 1.47$ for $q = 2$ (f), $q = 3$ (g), $q = 4$ (h)). We observe that the oscillations persist in time independently on the value of $q$ confirming that the BTC phase is not affected by the interactions along the $\x$ direction.}
    \label{Fig:pp_p2_q}
\end{figure*}
Actually, this is not valid in the case $p = 3$ because the trajectories surrounding the paramagnetic solutions, as shown in panel (c), are very slowly dampedtoward a time-independent value.

In the bottom panels of Fig.~\ref{Fig:pp_p3_p4} we plot the trajectories of the longitudinal magnetization, that in the phase portrait are highlited in blue and red on the Bloch sphere. From these figures emerge that, while in the case $p = 4$ (d) we have a closed orbit, for $p = 3$ (c) the trajectories are attracted toward the ferromagnetic solution with damping (which can be very slow, depending on the initial conditions) that eventually suppresses the oscillations, ruling out the possibility of BTCs.
An analysis of the oscillations amplitude $A(t)$ shows, indeed, that for $p>1$ it decreases in time with a $p$-dependent power law.
When $q=1$ the scaling is
\begin{equation}
    A(t)=B\ t^{-\frac{1}{p-1}} \ ,
\end{equation}

where the precise value of $B$ depends on the other system parameters.

Let us now consider the case of a $\Ztwo$ preserving Hamiltonian by fixing $p = 2$. In the top panels of Fig.~\ref{Fig:pp_p2_q} we plot the phase portraits for $q = 1, 2, 3, 4$. 
Although the complexity of the phase portrait increases by increasing $q$, there are always some periodic orbits signaling the presence of BTCs. The blue and the orange diamonds in panel (a) Fig. \ref{Fig:pp_p2_q} mark the parameters used in Fig. \ref{fig:exact_diff}.
The bottom panels show the trajectories in time of the longitudinal magnetization evaluated over the red trajectories, confirming the persistence of the oscillations and the emerging of a time-crystal order.

Although we do not have rigorous proof, our numerical analysis strongly suggests our claim on the symmetry requirement: if the dissipation does not break the discrete symmetry of the Hamiltonian, closed periodic orbits do not appear.

To conclude this section, we remark that the presence of the BTC phase, regardless of the value of $q$, has an important implication on the robustness of the phase itself.
When $q=1$, the $x$-term in the Hamiltonian can be interpreted as a transverse field or coherent driving. In this case, any fluctuations on its direction could break the $\Ztwo$ symmetry and inevitably destroy the BTC phase.
If $q=p=2$, instead, the Hamiltonian contains only interaction terms and no external field or driving. 
In this case the interactions concur to {\em stabilize} the BTCs phase since the symmetry $\Ztwo \times \Ztwo$ cannot be broken by noise in the Hamiltonian parameters.

\subsection{Finite-$N$ analysis}\label{sec:exact}

To corroborate the mean-field analysis done so far, 
we solve the exact quantum dynamics in finite systems and present three indicators that identify the BTC phase:
(i) the oscillations of the expectation value of the magnetization and the link between the damping rate and the system size $N$,
 (ii) the part of the Liouvillian spectrum closest to the origin of the complex plane, (iii) the structure of the stationary state density matrix $\rhoss$.

\subsubsection{Oscillations of the magnetization}\label{sec:mag}

Since the Lindblad equation conserves the total spin, we can restrict the analysis to the sector of the Hilbert space with total spin $S^2 = \frac{N}{2}\left(\frac{N}{2}+1\right)$. This allows us to access numerically systems of the order $N\sim 10^2-10^3$, depending on whether we are interested in the Liouvillian spectrum or only in the time evolution.
All numerical results have been obtained using the QuTiP~\cite{qutip1,qutip2} Python package.
In what follows, we mainly focus our attention on the Hamiltonian in Eq.~\eqref{Eq:Hamiltonian_z}, corresponding to the choice $p=2$, $q=1$.

\begin{figure}
    \centering
    \includegraphics[width=8cm]{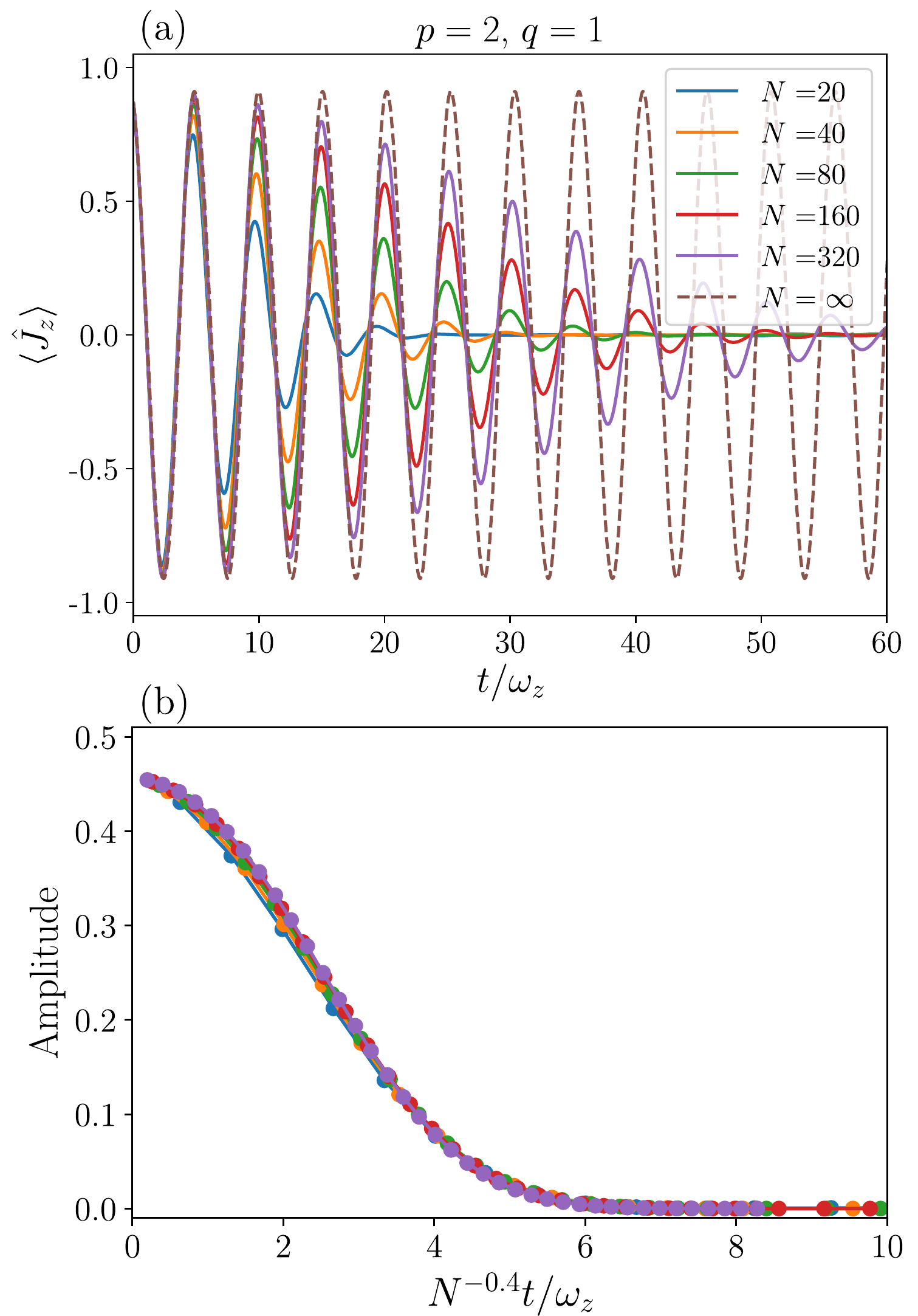}
    \caption{(a): Magnetization $\langle \Jz\rangle$ versus time $t$ in the BTC phase, for several values of the system size $N$ and interaction ranks $p=2$, $q=1$.
    The dashed line corresponds to the trajectory in the thermodynamic limit, obtained through the mean-field equations \eqref{Eq:dynamics}.
    The other parameters of the model are $\omega_x=3\omega_z$ and $\delta\Gamma = 0.2 \omega_z$.
    (b): Oscillations amplitude as a function of the rescaled time $\tilde{t}_N = t N^{-0.4}$. The nice collapse of the different curves suggests that the decay time has a power-law divergence $N^{0.4}$. }
    \label{fig:exact_mzvst}
\end{figure}

First, let us consider the finite-$N$ dynamics in the BTC phase.
Fig.~\ref{fig:exact_mzvst}(a) shows the main features of a typical trajectory of the magnetization in the BTCs phase.
Differently from the mean-field case, the magnetization presents oscillations decaying with a damping strength that decreases for increasing system size $N$, which finally vanishes in the thermodynamic limit.
By performing a data collapse of the oscillation amplitudes, we found that the decay rate follows a power-law compatible with $N^{-0.4}$, as can be observed in Fig.~\ref{fig:exact_mzvst}(b). 
This exponent, however, is non-universal and it depends on $p$ and $q$. 
For instance, in the free spin case in Eq.~\eqref{Eq:Hamiltonian_free} with $\omega_z=0$ studied in detail in Ref.~\cite{Iemini17}, the decay rate decreases as $N^{-1}$.

The period of the oscillations, instead, is practically size-independent and, in fact, the Fourier spectrum is always peaked around the same frequency with a small broadening that reduces by increasing $N$.  
In the BTC phase, this behavior is independent of the initial condition and, qualitatively, it holds $\forall p$ (even).

\begin{figure}
    \centering
    \includegraphics[width=8cm]{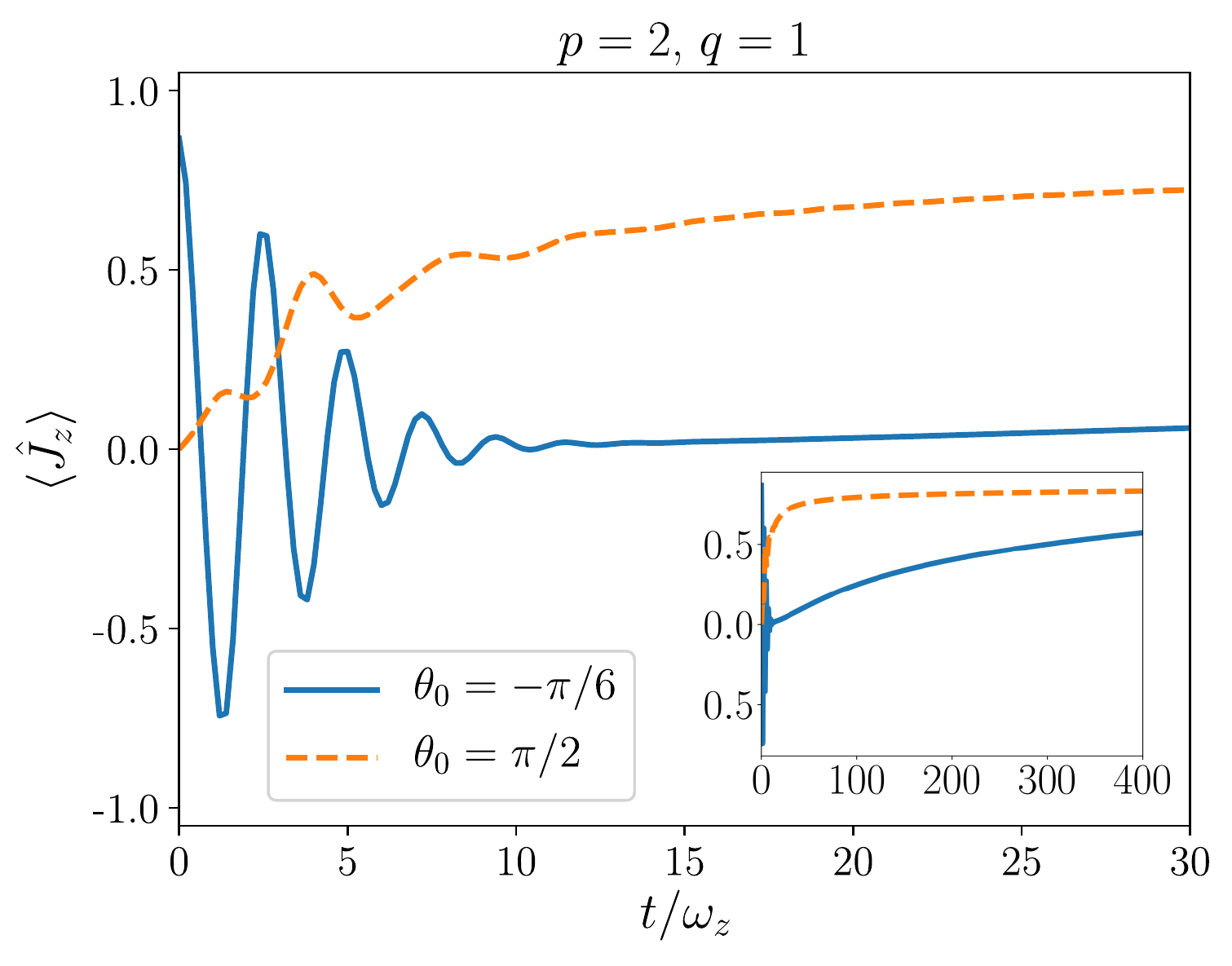}
    \caption{Comparison of the time dependence of the average magnetization for two different initial conditions in the coexistence phase: all spins aligned along the direction $(\theta_0,\phi_0)=(-\frac{\pi}{6},0)$ (solid blue curve) and $(\theta_0,\phi_0)=(\frac{\pi}{2},0)$ (dashed orange curve).
    The system size is $N=100$ and the other Hamiltonian parameters are $p=2$, $q=1$, $\omega_x=\omega_z$. The dissipation is $\delta\Gamma=0.1\omega_z$.
    The inset shows the same date on a longer time scale, where one can see the slow relaxation of the trajectory corresponding to the boundary time crystal towards the ferromagnetic stationary state.}
    \label{fig:exact_diff}
\end{figure}

In the coexistence phase (BTC+F), periodic orbits and relaxation dynamics coexist in the thermodynamic limit.
When one focuses on finite-size effects, this behavior persists for a finite-time window, as shown in Fig.~\ref{fig:exact_diff}.
Depending on the initial conditions, the trajectory might display both $N$-dependent damped oscillations or a size-independent relaxation toward the ferromagnetic stationary state. 
Eventually, this apparent bistability breaks after a time scale $\tau_N \sim N$, when the trajectory deviates from the time-crystal density matrix to slowly approach the unique steady state (inset of Fig.~\ref{fig:exact_diff}).
The relative Liuvillian spectrum is qualitatively similar to that of the BTC phase with a stationary state that is ferromagnetic instead of paramagnetic.

\subsubsection{Liouvillian spectrum}\label{sec:spec}
\begin{figure}
    \centering
    \includegraphics[width=8.5cm]{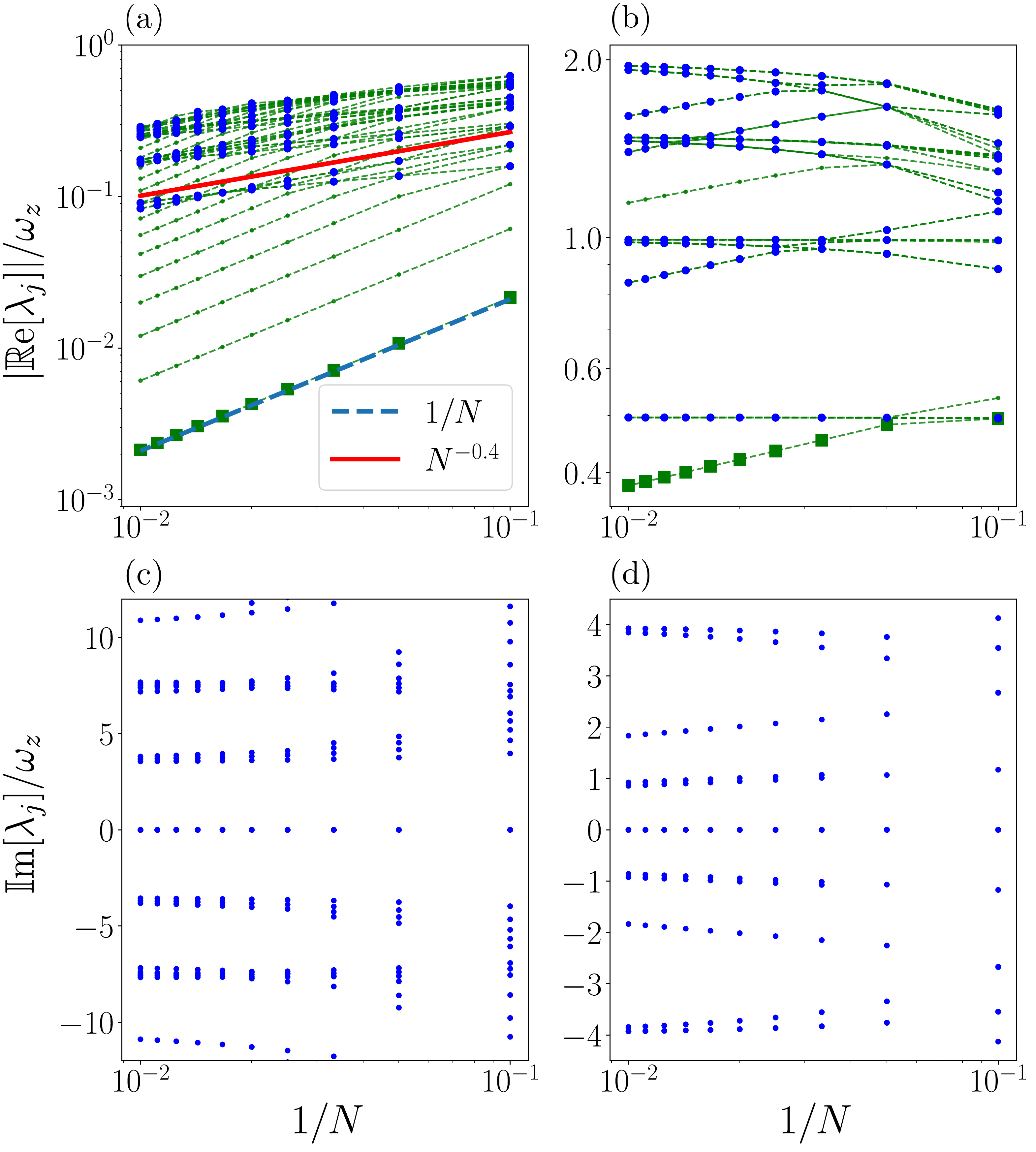}
    \caption{Real and imaginary parts vs $1/N$ of the 21 eigenvalues of the Liouvillian with $\Re{\lambda^{(N)}_i}$ closest to zero.
    The large blue circles in the upper panels highlight eigenvalues with a nonzero imaginary part.
    The dashed blue line in (a) and highlights the $1/N$ scaling of the Liouvillian gap (green squares); the solid red line shows the possible $N^{-0.4}$ scaling of the eigenvalues with nonzero imaginary parts.
    (a) and (c): BTC phase with $p=2$, $q=1$, $\omega_x=3\omega_z$, and $\delta\Gamma=0.2\omega_z$.
    (b) and (d): ferromagnetic phase with $p=2$, $q=1$, $\omega_x=0.25\omega_z$, and $\delta\Gamma=0.5\omega_z$.}
    \label{fig:Liuv_spec}
\end{figure}

The properties depicted above are reflected in the eigenspectrum of the Liouvillian. In Fig.~\ref{fig:Liuv_spec} we show the behavior of the real and the imaginary part (top and bottom panels, respectively) of the eigenvalues as a function of the system size $N$ for the BTCs phase, in (a) and (c), and the ferromagnetic phase in (b) and (d).

In the BTC phase, see Fig.~\ref{fig:Liuv_spec}(a), the real part of the spectrum roughly divides into a set of real eigenvalues that decreases as $\Re{\lambda^{(N)}} \sim N^{-1}$, and a set of complex eigenvalues associated with the BTCs (blue circles) with a much slower decay.
Again, the only exception is the free spin case in which, as mentioned, we find a power-law behavior.
Because of the computational effort needed to diagonalize the Liouvillian superoperator, it is hard to extract a precise scaling; however, we found that its behavior is compatible with the power-law decay $N^{-0.4}$, in agreement with the damped oscillations shown in Fig.~\ref{fig:exact_mzvst}.

The imaginary part in Fig.~\ref{fig:Liuv_spec}(c), instead, clearly saturates to values independent from $N$. Surprisingly, the oscillation frequency of the magnetization is not given by the ``quantization'' of  the imaginary part of the spectrum, in contrast to to what happens in the free case.
In absence of interactions, i.e. $p=0$, $q=1$, the frequency is indeed independent on the initial condition and can be extracted directly from the Liouvillian~\cite{Iemini17}.
In the present case, instead, the $\Jz^p$ interaction introduces a dependence of the frequency on the initial conditions, which might also be due to the presence of Liouvillian eigenstates with imaginary parts that are not perfectly commensurate.

In (b) and (d), we plot the Liouvillian eigenvalues for the ferromagnetic phase.
In this case, the behavior of the real part of the eigenvalues is markedly different from the previous one. 
There is still at least one real eigenvalue decreasing as $N^{-1}$, suggesting the possibility of multiple steady-states in the thermodynamic limit. Despite this, the Louvillian gap is much larger than the corresponding one in the BTC phase, as one can appreciate from the different scales of the two vertical axes. 
More importantly, the real part of the lowest complex eigenvalues (blue circles) do not display any dependence on $N$, which means that any oscillatory behavior is doomed to decay also in the thermodynamic limit.
\subsubsection{Steady-state density matrix}\label{sec:rhoss}
\begin{figure}
    \centering
    \includegraphics[width=0.49\textwidth]{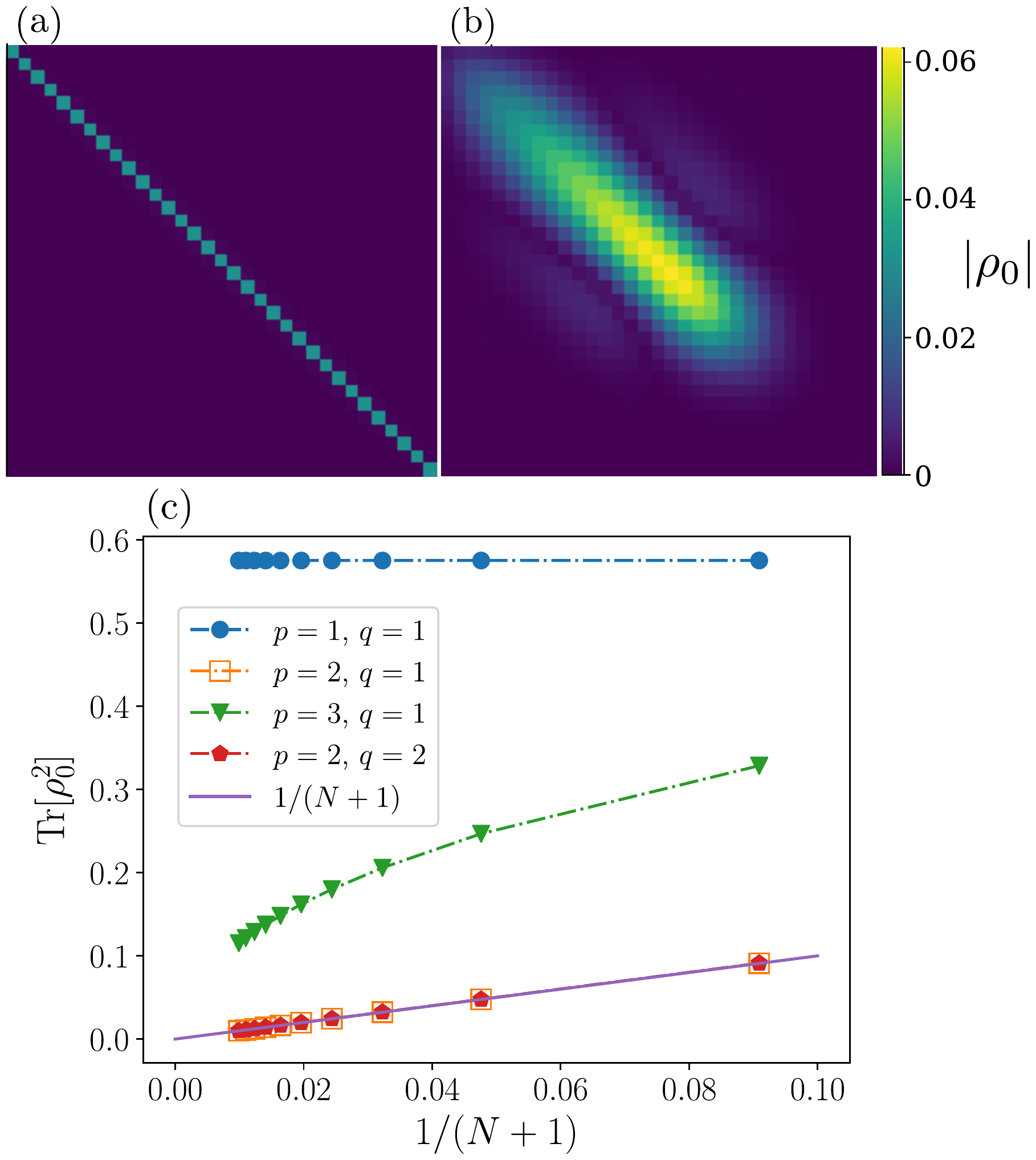}
    \caption{(a) and (b): absolute value of the stationary-state density matrix for $N=30$, $q=1$, $\omega_x=3\omega_z$ and $\delta\Gamma=0.2\omega_z$. In (a) $p=2$, while in (b) $p=3$.
    The density matrix elements are computed in the basis of the $z$-magnetization $\Sz=\frac{N}{2}\Jz$.
    (c): purity of the stationary-state density matrix $\Tr{\rho_0^2}$ versus $1/(N+1)$ for several values of $p$ and $q$. When the BTC phase is present, the purity is very close to the lowest possible in the subspace of the Hilbert space where the dynamics is constrained, namely $(N+1)^{-1}$. 
    When the system relaxes towards a state with finite $z$-magnetization, instead, the purity is much larger.
    }
    \label{fig:rhoSS}
\end{figure}

Finally, we would like to briefly comment on the stationary state density matrix $\rhoss$.
When the system is in the BTC phase, the density matrix $\rhoss$ of the stationary state for finite size $N$ is independent on the choice of individual parameters and acquires the characteristic structure shown in Fig.~\ref{fig:rhoSS}(a).
Besides some corrections of the order $O(N^{-2})$, $\rhoss$ is proportional to the identity matrix, where all eigenstates of $\Sz$ are populated uniformly.
This suggests that there is an emergent $\Zn$ symmetry, generated by a unitary ladder operator $\hat{Q}$ that acts on the eigenstates $\ket{m}$ of $\Sz$ as $\hat{Q}\ket{m}=\ket{m+1}$.

At finite size $N$ this symmetry is not exact, but it gets closer to an actual $\Zn$ invariance for increasing values of $N$, until it becomes a U$(1)$ symmetry in the thermodynamic limit.
A similar mechanism to simulate a U(1) lattice gauge theory has been proposed in Ref.~\cite{Zohar_PRA2013}. 
In the present case, we argue that this U$(1)$ symmetry is associated to the closed orbits arising in the mean-field description of the thermodynamic limit,
where the dynamics keeps the memory of the initial state, modulo a rotation. This result is further validated by the correspondence between the emergence of time crystals and dissipative phase transitions as shown in Ref.~\cite{Minganti_2020_arXiv}.

As a consequence of its structure, the BTC steady-state density matrix {\em cannot} be described with a mean-field approximation, 
meaning that it cannot be written as a direct product of single spin density matrices, 
since the fluctuations of the magnetization components {\em diverge} in the thermodynamic limit.
By a simple calculation, see App. \ref{App:A3}, it is possible to show that it does not exist a factorized density matrix $\rhoss$ giving $\Tr{\rhoss \Sx}\simeq\Tr{\rhoss \Sy}\simeq\Tr{\rhoss \Sz}\simeq 0$, while preserving $\hat{S}^2$.
It is more appropriate to think of the finite size steady state as the average of the magnetization vector over a BTC trajectory in the thermodynamic limit. 
This picture holds for any parameter choice where the BTC phase exists and it is in perfect agreement with the analysis of the semiclassical trajectories in Ref.~\cite{tcsavona}.

However, it is important to stress that this structure emerges independently in each eigenspace of the total spin $\hat{S}^2$.
Since each of these sectors of the Hilbert space has a different dimension, the full density matrix is
\begin{equation}
    \oprho_0 \simeq  \sum_{n=0}^{N/2} \left[ \frac{g(n)}{2n+1} \sum_{m=-n}^n \ket{n,m}\bra{n,m} + O(n^{-2}) \right] \ ,
\end{equation}
where $n$ and $m$ are the quantum numbers associated with $\hat{S}^2$ and $\Sz$ respectively, and we assumed the number of spin variables $N$ to be even. Above,
$g(n)$ is the degeneracy of the corresponding subspace.
Note also that the corrections to the identity-like structure becomes more and more relevant the smaller the total magnetization $n$ is.  
Hence, $\rho_0$ is not at all close to the identity in the full Hilbert space but only if we look at a fixed (large) value of of $n$, as we did in our analysis with $n=N/2$, corresponding to the maximally polarized subsector. 

%
%

In the ferromagnetic and mixed phases, the stationary-state density matrix resembles, instead, a coherent spin state, with a degree of spin squeezing \cite{Ma_2011} which depends on the values of the Hamiltonian parameters, $p$ and $\delta\Gamma$ in particular.
As an example, in Fig.~\ref{fig:rhoSS}(b) we show the absolute value of $\rhoss$ for $p=3$.
It is evident that, differently from the BTC steady state, the steady state also displays appreciable coherences. This feature provides further interesting elements on the generality of the model of Eq.~(\ref{Eq:Hamiltonian_generic}) also with regard to the steady-state properties of the driven-dissipative system, e.g., for subradiant and subradiance, coherence and spin squeezing \cite{DallaTorre16,Munoz_2019}.

To compare quantitatively the two cases, in Fig.~\ref{fig:rhoSS}(c) we plot the purity of the density matrix $\Tr{\rhoss^2}$ as a function of the system size $N$.
In the BTC phase, corresponding to $p$ even, the purity is practically the lowest possible $(N+1)^{-1}$, the same as the identity matrix. 
$\rhoss$ is indeed a mixed state without coherences which satisfies $\Tr{\rhoss \Sx}\simeq\Tr{\rhoss \Sy}\simeq\Tr{\rhoss \Sz}\simeq 0$.
For odd values of $p$ the purity is much higher, showing that the steady-state density matrix is quantitatively and qualitatively different in the two phases.

\section{BTCs existence condition 2: Collective dissipation
}\label{sec:colldecay}
\begin{figure}
    \includegraphics[width = 0.49\textwidth]{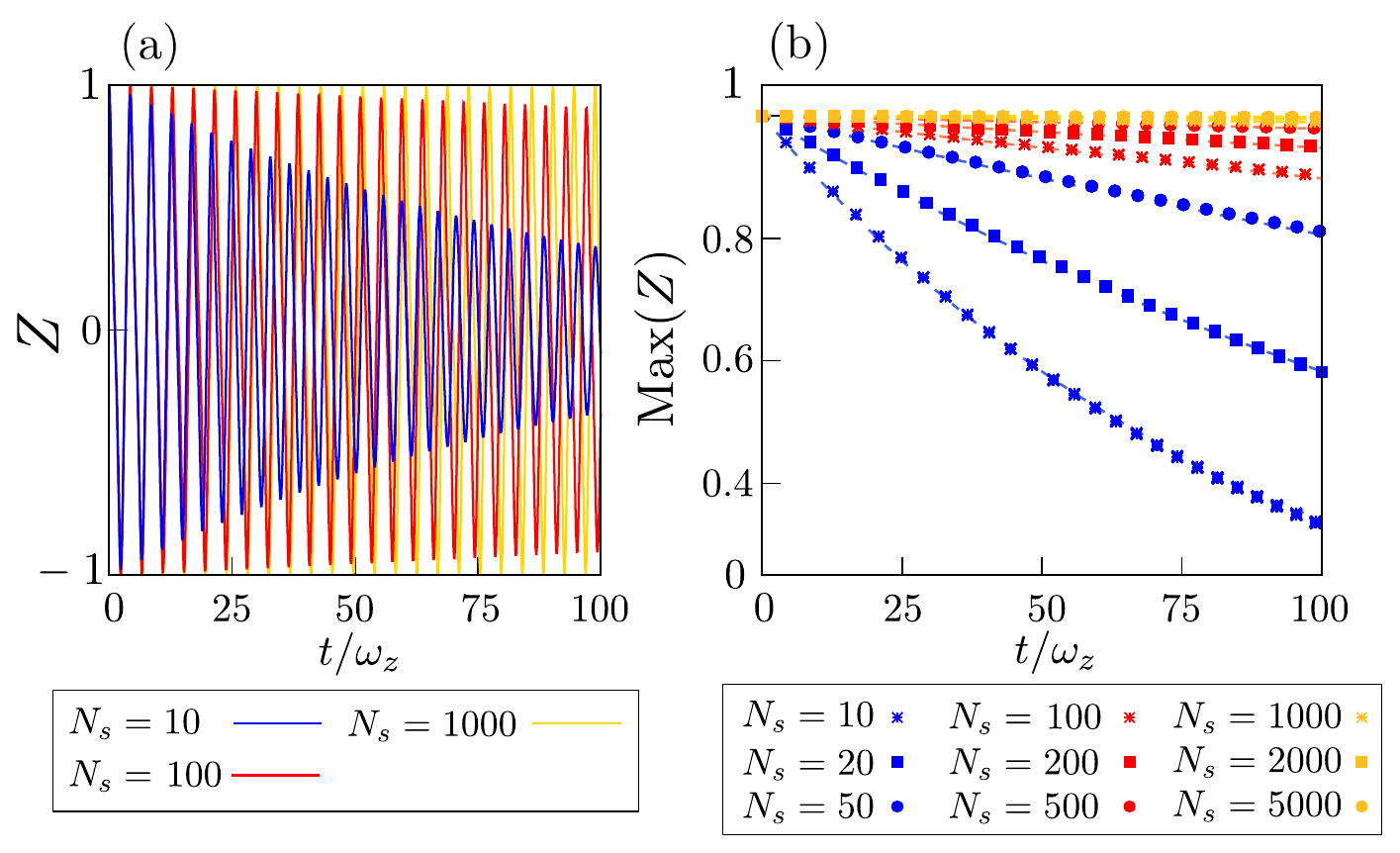}
    \caption{(a): Trajectories of the longitudinal magnetization $Z$ as a function of time for $N_s = 10, 100, 1000$, fixed $\dg = 0.2\omega_z$ and $\omega_x = 1.1\omega_z$. 
    The dashed line is the benchmark case with global dissipation processes $N_s = N$. 
    Reducing the value of $N_s$ the trajectories are affected by dephasing and damping effects that are as stronger as smaller $N_s$. 
    (b): Oscillation amplitude of the $Z$ component of the magnetization (full squares) as a function of time, for different values of $N_s$, and  $\dg = 0.2\omega_z$ and $\omega_x = 1.1\omega_z$. 
    The data are fitted with functions parametrized as $f(t)= \exp \left(- \beta t/N_s \right)$ (dashed lines), with the fit parameter being $\beta = 0.11$.}
    \label{Fig:max_damp}
\end{figure}

In this section, we show that only in the presence of collective dissipation processes only, the dynamics of Eq.~(\ref{Eq:dynamics}) displays time crystallinity. To this purpose, we follow the dynamics of the simplest time crystal Hamiltonian described in Eq.~\eqref{Eq:Hamiltonian_z}, i.e. when $p=2$ and $q=1$, coupled to a bath through a string of jump operators $J^\pm_s = \sum_i^{N_s} \sigma_i^\pm/N_s$, with $N_s \le N$.
Note that these Lindblad operators do not conserve the total angular momentum $J^2$. 
The mean-field equations of motion read
\begin{equation}
\left\{
\begin{aligned}
	& \dot{X} = 2p \omega_z  Z^{p-1}Y - 2\left(\dg Z_s +\frac{\overline{\Gamma}}{N_s}\right) X_s, \\
	& \dot{Y} = 2XZ \left (q\omega_x X^{q-2}- p\omega_z Z^{p-2} \right) - 2\left(\dg Z_s +\frac{\overline{\Gamma}}{N_s}\right) Y_s, \\
	& \dot{Z} = -2q\omega_x Y X^{q-1} + 2 \dg \left(1 - Z^2 + \frac{1}{N_s}\right) -\frac{2\overline{\Gamma}}{N_s} Z_s,
	\label{Eq:dynamics_Ns}
\end{aligned}\right.
\end{equation}
where we have defined $\overline{\Gamma} = \Gamma_\uparrow + \Gamma_\downarrow$ and $\alpha_s = \braket{J_\alpha^s}$.  The results in Sec. \ref{Sec:BTCs} are recovered in the thermodynamic limit when $N_s = N$. 
In (a) of Fig.~\ref{Fig:max_damp} we show the trajectories of the magnetization $Z$ as a function of time for three different values of $N_s = 10, 20, 50$; with $\dg = 0.2\omega_z$ and $\omega_x = 1.1 \omega_z$. By comparing with the collective dissipation trajectory (gray dashed line) it emerges a twofold effect, both on the phase and on the amplitude of the oscillations, due to the finite range of the dissipation operators. 
The trajectories for different values of $N_s$ have slightly different periods, leading to a progressive dephasing which, however, has no relevant effect on the time-order.
The more relevant consequence of the finite range of the bath operators is the damping that reduces the amplitude of the oscillations, eventually recovering a time-translational invariant state.

To analyze the $N_s$ dependence of the time scale over which time order is destroyed, 
in (b) of Fig. \ref{Fig:max_damp} we plot the amplitude of the oscillations of $Z$ as a function of time, for different $N_s$. 
The dashed lines, obtained by fitting the data, behave as
\begin{equation}
    f(t) = \exp \left(- \beta t/N_s \right)\ ,
\end{equation}
with $\beta = 0.11$. 
This means that the oscillations are exponentially damped in time, with a rate which decreases as a power law of the length of the string of the dissipation operators. 
Hence, only in the limit of global dissipation processes, $N_s \to N$, the oscillations persist for arbitrarily long time. 
This result confirms the intuition by Riera-Campeny and collaborators~\cite{Riera-Campeny2020} that global dissipation processes are a key ingredient to observe the boundary time crystal.
Our intuition is that this is an application of a more general condition that requires at least a conserved quantity in the dynamics. If so, in fact, the system cannot loose the information on the initial state during the evolution and is prevented to relax toward a time-independent state.  

\section{Physical interpretation}

In our investigation, we relied only on numerical evidences to support our claims, while rigorous proofs are left for future studies.
However, we conjecture some physical interpretations and connections with discrete time crystals to help reaching a more complete understanding of the phenomenon. 

Of the two conditions, the claim that the Hamiltonian should be $\Ztwo$ invariant and the dissipation should break this symmetry is the harder to physically justify.
The role of the symmetry in the Hamiltonian might be similar to that played in Floquet time crystals.
There, the breaking of discrete time translation usually comes from a periodic driving that at each period connects different symmetry-broken sectors of the system's $\mathbb{Z}_n$ symmetric Hamiltonian ~\cite{tc7}.
Similarly, we observe that BTC trajectories oscillate between two finite values of the magnetization, i.e. between two states that break the Hamiltonian $\Ztwo$ symmetry.
If there is indeed a connection with Floquet time crystals, this suggests that the BTC phase can be observed when the Hamiltonian has a general discrete symmetry.

A characteristic behavior for even $p$, as explained in Sec. \ref{sec:symmetry}, is that the operators $\Jp$ and $\Jm$ play alternatively the role of an incoherent dissipation and driving, depending on the sign of the magnetization. 
This continuous switching of the roles of the Lindbald operators concurs to stabilize the periodic orbits, which otherwise collapse to a stationary state.
However, since the presence of an additional symmetry along the $x$ direction does not preclude the emergence of BTCs, our first condition can probably be generalized as the condition for the Hamiltonian to have a discrete symmetry group which is reduced to a smaller one when the system is coupled with the environment.

From a more mathematical perspective, it is known that dissipation can lead to nontrivial asymptotic states in the presence of degeneracies in the Hamiltonian spectrum \cite{Harbola2006, Cattaneo2020, Dorn2021}. 

The second condition we address is that we need the jump operators to act collectively on the whole system, i.e. to have the form $J^\pm = \sum_i^N \sigma_i^\pm /N$. 
In this way, the Lindblad operators commute with the total angular momentum $\hat{J}^2$, which is a strong symmetry of the complete model and, consequently, a conserved quantity~\cite{Albert_2014_PRA}.
Thus, in the present paper, the condition of collective pumping/dissipation and the conservation of $\hat{J}^2$ are used interchangeably. 

In a more general framework,
a conservation law reduces the effective dimension of the space in which the dynamics takes place.
In the BTC phase, this reduction prevents the system to loose memory on its initial conditions and to relax toward a stationary state, allowing the emergence of periodic trajectories. 
We believe that the key element here is the presence of a (quasi)conserved quantity during the open dynamics, independently of it being also a strong symmetry of the Lindblad equation, thus leading to a weaker constraint.
This can be observed in spin models with power-law decaying interactions, where $\hat{J}^2$ does not commute with the Hamiltonian but the system might be still rigid enough to observe the emergence of BTCs.


\section{Experimental realizations}\label{sec:experiments}
In this section, we discuss how to experimentally implement the proposed model. We discuss two main aspects: relevant quantum simulation platforms and the engineering of higher-order spin interactions. 

We have shown that the spin algebra can be implemented in ensembles of spin-$\frac{1}{2}$ with all-to-all connectivity.
Rydberg atoms \cite{Henriet_2020}, trapped ions \cite{Zhang_2017}, artificial qubits in superconducting circuits \cite{Puri_2017}, and color defects in diamond \cite{Angerer_2018} are the most prominent platforms in which such all-to-all connectivity has been demonstrated. Let us point out that hybrid quantum systems \cite{Xiang_2013_RMP,Clerk_2020}, such as color vacancies in diamond coupled to a single superconducting resonator offer the possibility to smooth out a long-range photon-mediated interaction on several emitters \cite{Angerer_2018}. 

With regards to tuning the order of interactions, $p$, in the Hamiltonian of Eq.~(\ref{Eq:Hamiltonian_generic}), the task can be divided in the standard cavity QED interactions, $p,q\leq2$ and in the more exotic many-body interactions, $p,q>2$. In general, up to dipole approximation, the light-matter interaction that characterizes cavity QED systems involves only two particles per time, i.e. annihilation/creation of bosonic/fermionic particles. The interaction mediated by bosons can in several cases be traced out, leading to effective spin-spin interactions ($p=2$).  
Spin-spin interactions are also the outcome of several microscopic processes, such as genuine magnetic interaction, internal-external degrees of freedom coupling in ions, Van der Waals forces in neutral atoms and so on. 
Using the quantum simulation paradigm \cite{Georgescu_2014_RMP}, quantum phase transitions and dissipative phase transitions with long-range interactions have been engineered in trapped-ion and Rydberg-atom simulators \cite{Zhang_2017,Keesling_2019_Nature,Scholl_2020_arXiv,Ebadi_2020_arXiv}, although a power-law decay in the interaction length is common. 
In artificial atoms such as superconducting qubits, effective spin-spin interactions have been first proposed \cite{TsomokosNJP08} and more recently implemented \cite{Puri_2017,Xu_2020_ScienceAdv}, reproducing the physics of paradigmatic spin models, the Ising and Lipkin-Meshkov-Glick models.

The couplings for $p,q>2$ generally represent an indirect process and have been theoretically proposed in the context of spin glasses \cite{Kirkpatrick_1987_PRB}, which have become recently amenable to quantum simulation \cite{Roy_2019_PRA,Harris_2018_Science,Rotondo_15}. Direct interactions show up in the non-perturbative coupling between light and matter in the so-called ultrastrong and deep-strong coupling regimes \cite{Kockum_2019}, where it is not possible to arbitrarily tune higher-order coupling strengths, and where they occur in the boson-fermion framework. 
Other promising candidates to realize three-body interactions are 
hard-core bosons\cite{Pachos_PRL2004} or polar molecules in optical lattices~\cite{Buchler_NatPhys2007,Capogrosso_PRB2009,Bonnes_NJP2010}, and trapped Rydberg ions~\cite{Gambetta_PRL2020_1,Gambetta_PRL2020_2}, which represent the state-of-the-art for controlling complex many-body systems.

Recently, the multiple excited states of natural and artificial atoms have been exploited to implement quantum systems beyond qubit systems, but with qudits or in general bosonic degrees of freedom, e.g., molecules or transmons in superconducting devices. It is important to point out that spin models can be mapped to bosons (and viceversa), e.g., with a Holstein-Primakoff approximation, but with the limitation of being in the diluted regime, i.e. where nonlinear effects are negligible \cite{Shammah_2018}. 

With this regard, higher order interactions in bosonic systems based on superconducting circuit devices have recently been implemented, providing more freedom to tune higher-order interactions and suppress lower-order ones, exploiting nonlinearities in Josephson junctions \cite{Kockum_2019_Springer,Touzard_2018,Lescanne_2020,Vrajitoarea_2020_NaturePhys}, which can also be used to simulate spins with angular momentum greater than $\frac{1}{2}$ \cite{Neeley_2009_Science,Nori_2009_Science}.

To conclude this section, it is worth to mention that the all-to-all connectivity condition can probably be relaxed.
Restricting ourselves to $p,q \le 2$, it is a well established fact~\cite{Dutta_PRB2001,io3} that a long-range Ising model, with interactions between sites $i$ and $j$ decaying with distance as $J_{i,j}\propto |i-j|^{-\alpha}$, belongs to the same universality class of the fully-connected model considered in this paper when $\alpha<1$.
We believe that the ``rigidity'' of the collective spin dynamics arising from the long range interactions is sufficient to observe boundary time crystals in such systems, at least for sufficiently small values of $\alpha$. 
It is hard to anticipate, though, how the long time behavior changes if $\alpha \ge 1$ when fluctuations between different magnetization sectors may destroy the BTC phase. 

\section{Discussion and conclusion}
\label{sec:conclusions}
In this work, we discussed two conditions for the existence of boundary time-crystals in a large class of generalized $p$-interaction spin models with dissipation, showing that the emergence of this non-equilibrium phase is strongly related to the symmetries and the conserved quantities of the Hamiltonian and how they are affected by the dissipation. 

The first condition is that BTCs arise only if the spin Hamiltonian is $\Ztwo$-symmetric and the Lindblad operators explicitly break this symmetry. 
If so, the mean-field trajectories of the magnetization display persistent oscillations in time.
This result resolves the apparent contradiction between Ref.~\cite{Iemini17} and Ref.~\cite{Wang2020}: in the latter the symmetry condition is not satisfied, hence the absence of BTCs.

This reflects an emerging U$(1)$ symmetry of the equation of motion: closed orbits are invariant under rotations along the trajectory itself, meaning that the system preserves only partial information on the initial condition since all points on the same orbit lead to the same time-crystal dynamics.
At finite-size $N$, instead, the oscillations are affected by a damping rate that decreases by increasing $N$.
The U$(1)$ symmetry of the thermodynamic limit now seems to appear as an {\em approximated} discrete $\Zn$ symmetry, similar to that argued in Ref.~\cite{Zohar_PRA2013}, as suggested by the characteristic profile of the stationary-state density matrix $\rhoss$ associated to the BTC phase.
This condition provides a nice parallelism between boundary and discrete time crystals, a phase in which the subharmonic response is due to the exploration of different subsectors of some symmetric Hamiltonian or evolution operator~\cite{tc12,tc4,tc5,tc6}.

The second condition regards the operators coupling the system with the bath:
the time-crystal order is destroyed by dissipation processes that do not conserve the total angular momentum. 
In particular, we showed that if only a portion $N_s < N$ of the system is coupled to the same external bath, the amplitude of oscillations decreases exponentially with $N_s$. 

This result suggests that, more generally, BTCs arise only when there is at least one strong symmetry in the dynamics that prevents the system to lose information on the initial conditions and to attain a time-independent steady state. This draws an interesting parallelism with results that have shown how the symmetry sector to which an initial state belongs to can determine the current flows and quantum transport properties of the steady state \cite{Manzano_PRB2014}.
However, we do not exclude that other kind of long range couplings between the system and the environment may still allow for a stable BTC phase. 
This might be an interesting route to pursue to better understand if the robustness of the BTC phase is determined by the correlation length of the jump operators or strictly requires a homogeneous collective bath coupling.

Local dissipation induces also a dephasing in the mean-field trajectories, on top of the usual damping. 
This dephasing is probably due to the exploration of sectors with different total angular momentum; in the Dicke representation of permutational-invariant systems \cite{Shammah_2018}, this is equivalent to jumps between different Dicke ladders.
Since different sectors have different Liouvillian spectra, the difference between the imaginary parts of the eigenvalues might indeed be responsible for the progressive dephasing of the oscillations.
Understanding this mechanism could be an important step toward the full uptake of the BTC phenomenon. 
Moreover, it could give some hints on possible configurations in which the competition between global and local dissipation processes could give rise to interesting non-trivial phases of matters.

It is important to observe that the two conditions must be met in order to enter a BTC phase in the thermodynamic limit. A different, but similar phenomenology is that of decoherence-free subspaces \cite{Baumgartner_2008_NJP,Buca_2012,Albert_2014_PRA,Lieu_2020}, where time-dependent oscillations for long times are present at any system size, as investigated in the context of artificial giant atoms \cite{Baumgartner_2008_NJP,Kockum_2018,Buca_2019_NatureComm}.  
An example for the considered model is obtained for $p=0$, hence $\hat{H}\propto \Jx^q$ and a Lindblad jump operator equal to $\Jx$ (which could be interpreted as a stochastic jump or global dephasing), or another jump operator commuting with the Hamiltonian, such as spin-squeezing dissipation, $\Jx^2$, or a collective depolarizing channel.

These models preserve the $\Ztwo$ symmetry but are characterized by persistent oscillations at any system size $N$.
Indeed, the Liouvillian spectrum  at finite $N$ is qualitatively independent from the coupling strength with the bath, indicating that collective dephasing does not induce different phases in our model and acts somehow ``trivially'' on the system.

All the results of this work have been derived for a general class of dissipative spin models, which can be implemented in fully connected spin-$\frac{1}{2}$ ensembles, with a specific choice of the Lindblad dissipators, which can be engineered in quantum simulators, e.g., in superconducting-circuit-based quantum devices.
However, the fact that BTCs are clearly associated with symmetry properties and not with the specific form of the Hamiltonian suggest that these criteria might apply for a wider class of systems and Lindblad operators. 
We anticipate that one of the most interesting directions for future investigations would be to explore the universality of our results, aiming for a general formulation of the existence conditions of BTCs in terms of symmetries and conserved quantities that may apply to a generic quantum many-body system.

\acknowledgements 
The authors acknowledge useful discussions with M. Burrello, M. Dalmonte, R. Fazio, F. Minganti, D. Rossini, G.E. Santoro, A. Silva and F.M. Surace. G.P. thanks Giovanni Piccitto for useful comments. 
F.N. is supported in part by: Nippon Telegraph and Telephone Corporation (NTT) Research, the Japan Science and Technology Agency (JST) [via the Quantum Leap Flagship Program (Q-LEAP), the Moonshot R\&D Grant Number JPMJMS2061, and the Centers of Research Excellence in Science and Technology (CREST) Grant No. JPMJCR1676], the Japan Society for the Promotion of Science (JSPS) [via the Grants-in-Aid for Scientific Research (KAKENHI) Grant No. JP20H00134 and the JSPS–RFBR Grant No. JPJSBP120194828], the Army Research Office (ARO) (Grant No. W911NF-18-1-0358), the Asian Office of Aerospace Research and Development (AOARD) (via Grant No. FA2386-20-1-4069), and the Foundational Questions Institute Fund (FQXi) via Grant No. FQXi-IAF19-06.
M.W. is supported by the Villum Foundation (Research Grant No. 25310).
This project has received funding from the European Union’s Horizon 2020 research and innovation program under the Marie Sklodowska-Curie grant agreement No. 847523 ‘INTERACTIONS’


\appendix

\section{Derivation of the equations of motion}\label{App:A1}
In this section we derive the mean-field equations of motion of Eq.~\eqref{Eq:dynamics_Ns}. Let us assume to have a dynamics described by the master equation
\begin{equation}\label{Eq:a1_dyn}
	\begin{aligned}
		\dot{\rho} = - &i [\hat{H}, \rho] + N\Gamma^\uparrow \left(J^+_s \rho J^-_s - \frac{1}{2}\{J^-_sJ^+_s, \rho\} \right)\\
		+ &N\Gamma^\downarrow \left(J^-_s \rho J^+_s - \frac{1}{2} \{J^+_s J^-_s, \rho\} \right)\\
		= & \dot{\rho}^c + \dot{\rho}^d.
	\end{aligned}
\end{equation}

This evolution has two different contributions, a term that accounts for the coherent evolution, $\dot{\rho}^c$, and a term that accounts for the incoherent one, $\dot{\rho}^d$. 
The coherent part is simply obtained by evaluating the commutator with the Hamiltonian leading to
\begin{equation}
    \begin{aligned}
        &\dot{J}_x^c = 2 \sum_{n =  0}^{p-1} J_z^{p-1-n} J_y J_z^n, \\
        &\dot{J}_y^c = -2 \sum_{n =  0}^{p-1} J_z^{p-1-n} J_x J_z^n + 2 g J_z, \\
        &\dot{J}_z^c = -2gJ_y. \\
    \end{aligned}
\end{equation}

We want to evaluate the incoherent contribution to the equations of motion for the expectation values of the spin operators. This is given by $\braket{\dot{J^\alpha}} = \text{Tr} \left( \dot{\rho}^d J^\alpha \right)$ leading to
\begin{equation}
	\begin{aligned}
		\braket{\dot{J_\alpha}} = - &N\Gamma^\uparrow\text{Tr} \left(J^+_s \rho J^-_s J_\alpha - \frac{1}{2}\{J^-_sJ^+_s, \rho\}J_\alpha  \right)\\
		+ &N\Gamma^\downarrow \text{Tr}\left(J^-_s \rho J^+_s J_\alpha - \frac{1}{2} \{J^+_s J^-_s, \rho\}J_\alpha \right).\\
\end{aligned}
\end{equation}
By manipulating the equations we obtain
\begin{equation}
	\begin{aligned}
	\braket{\dot{J_\alpha}}	= &\frac{Ni\dg}{2}\braket{ \left\{J^x_s,  \left[J_\alpha , J_s^y \right] \right\}- \left\{J^y_s,  \left[J_\alpha , J_s^x \right] \right\} }\\
		+ &\frac{N\gb}{2}\braket{ \left[J^x_s,  \left[J_\alpha , J_s^x \right] \right]+ \left[J^y_s,  \left[J_\alpha , J_s^y \right] \right] }\\
		= &\frac{i\dg}{2N_s^2}\sum_{i,j}^{N_s} \sum_l^N \braket{ \left\{   \sigma^x_i,  \left[\sigma^\alpha_l , \sigma_j^y \right] \right\}- \left\{\sigma^y_i,  \left[\sigma^\alpha_l , \sigma_j^x \right] \right\} }\\
		+ &\frac{\gb}{2N_s^2}\sum_{i,j}^{N_s} \sum_l^N\braket{ \left[\sigma^x_i,  \left[\sigma^\alpha_l , \sigma_j^x \right] \right]+ \left[\sigma^y_i,  \left[\sigma^\alpha_l , \sigma_j^y \right] \right]}\\
		= -&\frac{\dg}{2N_s^2}\sum_{i,j}^{N_s} \braket{ \left\{\sigma^x_i, \epsilon_{\alpha y \beta} \sigma^\beta_j\right\}- \left\{\sigma^y_i, \epsilon_{\alpha x \beta} \sigma^\beta_j\right\} }\\
		- &\frac{\gb}{2N_s^2}\sum_{i,j}^{N_s} \braket{ \left[\sigma^x_i,  \epsilon_{\alpha x \beta} \sigma^\beta_j \right]+ \left[\sigma^y_i,  \epsilon_{\alpha y \beta} \sigma^\beta_j \right] },
	\end{aligned}
\end{equation}
where $\dg = \Gamma^\uparrow - \Gamma^\downarrow$ and $\overline{\Gamma} = \Gamma^\uparrow + \Gamma^\downarrow$.

\section{Derivation of the phase portrait} \label{App:A2}
In this appendix we provide some details on the derivation of the phase portrait. 
In general, the spin operators in the semi-classical approximation can be parametrized on the Bloch sphere by giving the radius $r$ of the sphere, the polar angle $\ph$ and the azimuthal one $\theta$. This mapping is singular in $\theta = 0, \pi$, therefore some care is needed in choosing the axis defining $\ph$ and $\theta$. The results of the paper have been derived by parametrizing the spin as
\begin{equation}\label{Eq:a1}
\left(X = r \sin \theta \cos \ph, \ Y = r \sin \theta \sin \ph, \  Z = r \cos \theta \right).
\end{equation}
The only exceptions are the results in Subsec. \ref{Subsec:x}, in which we assumed the polar angle to span the $yz$ plane (namely, we fixed $X = \cos \theta$). We will omit the derivation of this case since it can be obtained by simply extending the following calculations. 

Eq.~\eqref{Eq:a1} can be inverted leading to
\begin{equation}
    \begin{aligned}
        &r = \sqrt{X^2 + Y^2 + Z^2}, \\
        &\ph = \arctan(Y/X), \\
        &\theta = \arctan((X^2 + Y^2)/Z).
    \end{aligned}
\end{equation}
By deriving and substituting the expressions of $X, Y, Z$ in terms of $r, \ph, \theta$ we obtain, for the collective decay, the following equations of motion (since classically we have that $\ph$ and $\cos \theta$ are conjugate variables, it is better to derive the equations for $\cos(\theta)$ instead of $\theta$) 
\begin{equation}
\begin{aligned}
	&\dot{r} = 2 r \dg \cos \theta  \left(1 - \ r^2 \right), \\
	& \dot{\ph} =   -2 \omega_z p \cos^{p-1} \theta + 2 \omega_x q \cos \theta \sin^{q-2} \theta \cos^q \ph, \\
	&\dot{\cos \theta} = - \left(2 \omega_x \sin^q \theta \cos^{q-1} \ph \sin \ph + 2 \dg \left(1 - \cos^2 \theta \right) \right)/r. 
\end{aligned}
	\label{theta_phi_mf}
\end{equation}
We notice that $r = 1$ is a fixed point, hence the dynamics is constrained on the surface of the Bloch sphere.

\section{Mean-field description of the totally symmetric subsector}\label{App:A3}
Let us consider a generic single spin density matrix, parametrized as
\begin{equation}\label{eq:rho1}
    \oprho =\left[ \begin{array}{cc}
         a &  b \nep^{-i\phi}\\
          b\nep^{i \phi} & 1-a
    \end{array} \right]
\end{equation}
where $a$ and $b$ are both real numbers.
Its purity is defined as $\PP = \Tr{\oprho^2}=\frac{1}{2}\left[(2a-1)^2 + 1 + 4 b^2 \right] \leq 1 $. The reason why write it in this way will be clearer in the following.

We can construct a mean-field ansatz for the system by taking the tensor product of $N$ identical spins described by density matrices as Eq.~\eqref{eq:rho1}
\begin{equation}
    \oprho_{\rm mf} = \bigotimes_{j=1}^N \oprho_j \ .
\end{equation}

We can compute the expectation value of the total spin $\hat{S}^2=\Sx^2+ \Sy^2 + \Sz^2$ on this state.
First, let us rewrite it in terms of single-spin Pauli matrices
\begin{equation}
    \hat{S}^2 = \frac{3}{4}N + \frac{1}{4}\sum_{k, j\neq k }\Psigma^x_j\Psigma^x_k + \Psigma^y_j\Psigma^y_k+\Psigma^z_j\Psigma^z_k \ .
\end{equation}
Then, simple algebra leads to
\begin{equation}
    \Tr{\oprho_{\rm mf}\hat{S}^2} = \frac{3}{4}N + \frac{1}{4} N(N-1)\left[(2a-1)^2+4b^2\right] \ , 
\end{equation}
which can conveniently be rewritten in terms of the purity of the single-spin density matrix
\begin{equation}\label{eq:purity}
    \Tr{\oprho_{\rm mf}\hat{S}^2} = \frac{3}{4}N + \frac{1}{4} N(N-1) (2\PP-1) \ .
\end{equation}
If $\hat{S}^2$ is conserved and we consider in the maximally-polarized subsector, the total spin has to be equal to $\frac{N}{2}\left(\frac{N}{2}+1\right)$, which is compatible with Eq.~\eqref{eq:purity} only if $\PP=1$.
This means that a mean field ansatz for the density matrix {\em cannot} describe a mixed state in the maximally polarized subspace, but only pure ones. 
That is why to recover a mean-field approximation of mixed states such as that shown in Fig.~\ref{fig:rhoSS}, it is necessary to average over all magnetization vectors belonging to the same time-crystal trajectory.

\bibliography{biblio.bib,tesi.bib}

\begin{thebibliography}{131}
\bibitem [{not()}]{nota1}%
  \BibitemOpen
  \href@noop {} {}\bibinfo {note} {Throughout the paper, the word
  ``collective'' identifies an operator which is the sum over all sites of
  single spin objects, $\hat{O}=\sum_{j=1}^N \hat{o}_j$. Notice that $\hat{O}$
  is invariant under permutation of the spin indices.}\BibitemShut {Stop}%
\end{thebibliography}

\end{document}